\documentclass[11pt,a4paper]{article}
\usepackage{jheppub}

\usepackage{slashed}
\usepackage{xfrac}
\usepackage{subcaption}
\usepackage{booktabs}
\usepackage{array}
\usepackage{placeins}

\usepackage[colorlinks=true
,urlcolor=blue
,anchorcolor=blue
,citecolor=blue
,filecolor=blue
,linkcolor=blue
,menucolor=blue
,pagecolor=blue
]{hyperref}
\usepackage[all]{hypcap}

\newcommand{\plotwidth}{0.33}
\newcommand\subplotfeats[2]{
\begin{figure}[ht]
    \centering
    \vspace{-0.5\baselineskip}
    \begin{subfigure}[b]{\plotwidth\textwidth}
        \centering
        \includegraphics[width=\textwidth, page=4]{feature_hists_#1.pdf}
    \end{subfigure}
    \begin{subfigure}[b]{\plotwidth\textwidth}
        \centering
        \includegraphics[width=\textwidth, page=2]{feature_hists_#1.pdf}
    \end{subfigure}
    \begin{subfigure}[b]{\plotwidth\textwidth}
        \centering
        \includegraphics[width=\textwidth, page=5]{feature_hists_#1.pdf}
    \end{subfigure}
    \begin{subfigure}[b]{\plotwidth\textwidth}
        \centering
        \includegraphics[width=\textwidth, page=3]{feature_hists_#1.pdf}
    \end{subfigure}
    \begin{subfigure}[b]{\plotwidth\textwidth}
        \centering
        \includegraphics[width=\textwidth, page=6]{feature_hists_#1.pdf}
    \end{subfigure}
    \begin{subfigure}[b]{\plotwidth\textwidth}
        \centering
        \includegraphics[width=\textwidth, page=1]{feature_hists_#1.pdf}
    \end{subfigure}
    \vspace{-0.3\baselineskip}
    \caption[Feature distributions of #2 signal]{Distributions of vertex displacement $D_0/\gamma \beta_T$, vertex mass $m_{\rm vertex}$, vertex transverse momentum fraction $p_T^{\rm vertex} / p_T^{\rm jet}$, number of tracks associated to vertex $n_{\rm tracks}$, total number of jet constituents $n_{\rm obj}$, and total number of reconstructed displaced vertices $n_{\rm vertices}$ (not including primary vertex). If more than one vertex is reconstructed in a given jet the median value for vertex features is taken. The step-like features in the distribution of $p_T^{\rm vertex} / p_T^{\rm jet}$ are an artifact of requiring the sum of this variable over all jet vertices to be greater than 0.2. Jets with only one displaced vertex, which are the majority of background jets, are constrained to values greater than 0.2. Jets with two displaced vertices are constrained to a median greater than 0.1, etc.}
    \vspace{-8mm}
\end{figure}
}

\begin{document}
	
\title{Searching for dark jets with displaced vertices \\ using weakly supervised machine learning}

\author{Debjyoti Bardhan,\footnote{Present address: Department of Physics, Indian Institute of Science Education and Research, Pune, India.}}
\author{Yevgeny Kats,}
\author{Noam Wunch}
\affiliation{Department of Physics, Ben-Gurion University, Beer-Sheva 8410501, Israel}
\emailAdd{debjyoti.bardhan@acads.iiserpune.ac.in}
\emailAdd{katsye@bgu.ac.il}
\emailAdd{noamwunch1@gmail.com}

\abstract{If ``dark quarks'' from a confining hidden sector are produced at the LHC, they will shower and hadronize to dark sector hadrons, which may decay back to Standard Model particles within the detector, possibly resulting in a collimated spray of particles resembling a QCD jet. In this work we address scenarios in which dark hadrons decay with a measurable small displacement, such that the relevant background is dominated by heavy-flavor jets. Since dark sector parameters are largely unconstrained, and the precise properties of a dark QCD-like theory are difficult to compute or simulate reliably in any case, model-independent, data-based searches for such scenarios are desirable. We explore a search strategy employing weakly supervised machine learning to search for anomalous jets with displaced vertices. The method is tested on several toy signals, demonstrating the feasibility of such a search. Our approach has potential to outperform simple cut-based methods in some cases and has the advantage of being more model-independent.}

\maketitle

\section{Introduction}
\label{intro}

The diversity of particles and interactions in the Standard Model (SM), along with the multiple questions that the SM leaves unanswered, make it plausible for additional sectors of particles to exist in nature. One simple possibility is an exotic QCD-like sector~\cite{Strassler:2006im}. The fermions and gauge bosons of this sector can be charged under a new (``dark'') gauge group and neutral under the SM. They are termed dark quarks and dark gluons, in analogy with QCD. If the dark gauge group confines at low energies, the spectrum of this sector will contain composite states neutral under the new gauge group---dark hadrons. One motivation for these models is the possibility that dark hadron species that are stable on cosmological scales may account for dark matter~\cite{Bai:2013xga, Bernreuther:2019pfb, Beauchesne:2018myj, Beauchesne:2019ato}. However, such models are interesting also if they do not play this role.

If some portal (for example, a heavy mediator) couples the SM with the hidden sector, dark quarks can potentially be produced at the LHC. If dark quarks are produced, they will undergo parton showering and hadronization in the dark sector, similar to QCD quarks.  Species of dark hadrons that are stable on detector scales will escape the detector leaving a trail of missing energy. On the other hand, some species may be unstable, decaying back to the SM within the detector and forming a peculiar jet of SM particles. In this work we aim to obtain better sensitivity to such types of objects, known as \emph{dark jets}.

The collider signature of dark jets (see ref.~\cite{Albouy:2022cin} for a review) is greatly influenced by dark sector specifics. Many typical models would contain light dark pions $\pi'$, and dark vector mesons $\rho'$ and other hadrons with masses of order of the dark sector confinement scale $\Lambda_{\mathrm{QCD'}}$. In such scenarios, dark jets can be coarsely characterized by three parameters: the average fraction of momentum carried by stable (invisible) dark hadrons $r_{\mathrm{inv}}$, dark pion mass $m_{\pi'}$, and dark pion lifetime $\tau_{\pi'}$.

In cases with sizable $r_{\mathrm{inv}}$, a key signature will be missing energy $\slashed{E}_T$ aligned with a jet. This scenario of \emph{semivisible jets} was analyzed in refs.~\cite{Cohen:2015toa, Cohen:2017pzm}, where a search program for such models was proposed. References~\cite{Park:2017rfb, Cohen:2020afv, Kar:2020bws} suggested also making use of jet substructure variables. A search for resonant production of semivisible jet pairs was conducted by CMS in ref.~\cite{CMS:2021dzg}. It employed the cuts motivated by ref.~\cite{Cohen:2017pzm} to probe for models with intermediate $r_{\mathrm{inv}}$ and promptly decaying visible dark hadrons. This search also used a boosted decision tree with jet substructure inputs motivated by refs.~\cite{Park:2017rfb, Cohen:2020afv}. ATLAS has performed a search for nonresonant production of semivisible jets in ref.~\cite{ATLAS:2023swa}. Potential use of supervised deep neural networks (NN) for the classification of prompt, semivisible jets was studied in refs.~\cite{graphnet,Lu:2023gjk}, weakly supervised learning was considered in ref.~\cite{Finke:2022lsu}, and the use of an unsupervised NN, an autoencoder, was considered in ref.~\cite{Canelli:2021aps}.

Other types of dark jet scenarios, where missing energy is no longer a dominant signature, are also possible. For example, dark pions with macroscopic flight distances, $c\tau_{\pi'}$ of order $1-10$~cm, will manifest as highly displaced objects within the jet. A novel reconstruction object, Emerging Jet, has been proposed for the classification of such jets in ref.~\cite{Schwaller:2015gea}. A search for such objects, which are jets with few or no tracks originating from the primary vertex, was later conducted by CMS~\cite{CMS:2018bvr}. This search was sensitive to scenarios with large dark pion flight distance, where QCD background is scarce.

Overall, the large number of unknown dark sector parameters (gauge group, confinement scale, number of dark quark flavors and their masses, additional interactions within the dark sector, type of couplings to the SM and their strength), combined with the difficulty of simulating dark sector showering and hadronization reliably, call for model-independent and simulation-independent searches for anomalous jets. Machine learning (ML), and in particular weakly supervised ML, is a natural tool for such a task.

In this work we propose to employ weakly supervised ML for a largely model-independent, data-based search that would be sensitive to anomalous jets (such as dark jets) containing mildly displaced decays, so that the background is dominated by heavy-flavor jets. We will not assume the anomalous jets to contain missing energy since that case has already been explored a lot in the literature, but will instead rely on the presence of displaced objects. We will assume the anomalous jets to be pair produced in a decay of a heavy resonance.

The rest of the paper is organized as follows. In section~\ref{ML} we review the relevant ideas of weakly supervised machine learning. In section~\ref{proposed_search} we describe the proposed search, examine the most important backgrounds and define the features that will be made available to the NN. In section~\ref{benchmark_datasets} we define the datasets of signals and background that we use to simulate the search. We present the search simulation in section~\ref{example_search}. We discuss the results and state our conclusions in section~\ref{conclusions}. Appendix~\ref{evgen} describes the event generation. The details of the neural network classifier are provided in appendix~\ref{classifier_details}. Jet feature distributions of all benchmark signals compared with the background distributions are presented in appendix~\ref{feat_hists}. The fit procedure used for estimating bump significance is described in appendix~\ref{fit_procedure}.

\section{Weakly supervised machine learning}
\label{ML}

While the most traditional ML approach, that of \emph{fully supervised learning}, can provide very powerful classifiers, using it to search for physics beyond the Standard Model (BSM) requires specifying the exact BSM scenario that is being searched for (and being able to simulate it reliably). This makes fully supervised methods very model specific. In recent years, methods have been developed which lessen signal model dependence for selecting the test statistic. These methods provide different amounts of model independence, with the common trade-off of model independence vs.\ signal sensitivity.

An example of a completely model-independent test statistic is the output of an autoencoder trained on data (e.g., refs.~\cite{AE1, AE4, AE2, AE3}). This \emph{unsupervised learning} method, while being completely model agnostic, lacks sensitivity to many signals~\cite{Batson:2021agz,Collins:2021nxn}.

A more moderate approach, in the realm of \textit{weakly supervised learning}, requires knowledge of class proportions. In fully supervised training the true class of each training example, e.g.\ signal/background, is known and provided to the NN. Knowledge of class proportions means only knowing what fraction of training examples belong to each class. Using class proportions alone, a classifier can learn to distinguish between classes, while training directly on the mixed data. In ML literature this method goes by the name Learning from Label Proportions~\cite{yu2015learning, JMLR:v10:quadrianto09a}. It was shown to be effective in quark/gluon discrimination, where calculation of flavor proportions is possible~\cite{CWoLa1, Cohen:2017exh}.

This was extended to cases where label proportions are unknown, with the sole requirement of two event groups that have different signal proportions---this was termed \textit{Classification Without Labels (CWoLa)}~\cite{CWoLa2, CWoLa3}. To implement it in a search, one must separate the data to signal- and background-rich groups, based on some property of the signal model. In the case of a signal resonant in some parameter, the signal-rich sample can be obtained from selecting events near the resonance. This method goes by the name \textit{Extended Bumphunt}~\cite{CWoLa4} and was implemented by ATLAS in ref.~\cite{ATLAS:2020iwa}. Using CWoLa in a monojet search to enhance its sensitivity to semivisible jets was proposed in ref.~\cite{Finke:2022lsu}.

Another approach, \textit{Tag N' Train}~\cite{Amram:2020ykb}, suggests using signal dijet topology to obtain the mixed samples. This method uses \emph{cotraining}, with the dataset of dijet events being split into two \emph{views} of each event, one containing first-jet features and the other containing second-jet features. A classifier is trained to discriminate signal-like first jets from background-like first jets. A second classifier is similarly trained on second-jet features. Finally these classifier predictions are combined amounting to an event classifier. Each of the jet classifiers is trained using CWoLa, where signal- and background-rich labels are obtained from some criterion on the other jet in the event. In ref.~\cite{Amram:2020ykb} this criterion was taken to be a cut on the output of an autoencoder trained on the jet. It was shown in~\cite{Amram:2020ykb} that Tag N' Train and Extended Bumphunt can be effectively combined in searches for a dijet resonant signal. In the current work, we adapt this last approach to suggest a new search for dark jets.

\section{Proposed search}
\label{proposed_search}

We propose a largely model-independent, data-based search that would be sensitive to resonantly pair-produced anomalous jets containing mildly displaced decays.

\subsection{Strategy}
\label{search_strategy}

We first select for dijet events with displaced objects, as will be described in section~\ref{ev_select_section}. Next we define signal and background regions in dijet invariant mass based on a mediator mass and resonance width hypothesis.

An event classifier is obtained according to the following procedure. As in Tag N' Train~\cite{Amram:2020ykb}, each of the two leading jets in each event may be assigned a signal- or background-rich \emph{weak label} according to some condition on the ``other jet'' (among the two) in the event. In Tag N' Train, the ``other jet'' condition was based on an autoencoder output, which is fully signal model independent. We propose to use some model assumption, namely the fact that dark jets will often have more constituents than SM jets.\footnote{One could also design an analogous search that would be sensitive to scenarios in which BSM jets have fewer constituents than SM jets.} Therefore, we choose jet constituent count, $n_{\mathrm{obj}}$, as our weak classifier.\footnote{Other quantities, such as the number or the properties of the displaced vertices in the jet, could serve as alternative weak classifiers. Since our goal is to define a model-independent search, we want the criterion defining the weak labels to be simple and general and not optimized for any particular scenario. The more detailed use of the various features that might distinguish signal from background in each particular scenario is left to the neural network.} The two jets are ordered by descending $p_T$ and labeled $j_1$ and $j_2$. Signal-rich labels are assigned to jets within the signal region for which the other jet constituent count is greater than some chosen threshold $n^S_{\mathrm{obj}}$. Background-rich labels are assigned to jets coming from the entire mass range (signal region and sidebands) for which the other jet constituent count is smaller than some chosen threshold $n^B_{\mathrm{obj}}$. Using these S/B-rich labels, two classifiers are trained, one on $j_1$s and the other on $j_2$s. We define our test statistic for the event as a whole to be the product of the two jet classifier outputs. This quantity will tend to have higher values for signals than for background events. To avoid inference of events used for training, the data should be split into $k$-folds. The preceding steps should be repeated $k$ times, each time leaving a different fold out of training. The event classifier not trained on a given fold is used to classify the fold events.

The classifier is applied to both signal-region and sideband events, and a cut with efficiency $\epsilon_D$ for the data in that entire mass range is applied on the classifier output. The optimal value of $\epsilon_D$, i.e., most sensitive to signal, is model dependent and therefore several values should be used. After applying the cut, the invariant mass distribution in the sidebands is interpolated into the signal region. The expected event count in the signal region, based on the interpolation, is compared to the measured number of events in the signal region. The significance of the excess is estimated based on Poisson statistics and systematic uncertainties of the interpolation.

The search is to be conducted in the form of a bump hunt in dijet invariant mass, i.e.\ each mediator mass hypothesis, $m_{Z'}$, is considered separately. Resonance width can either be determined from simulation or scanned over (e.g., as in the \textsc{BumpHunter}~\cite{Choudalakis:2011qn}).

Once a significant excess is identified by the NN, the jets from the events that pass the NN cut can be examined manually to understand their nature.

\subsection{Event selection}
\label{ev_select_section}
Event selection for the proposed analysis is performed in two steps: a primary selection for dijet events adhering with trigger limitations, and a more tailored selection for events with displaced objects. Event selection requirements are summarised in table \ref{ev_req}. 

\subsubsection{Primary selection}
The main motivation for the primary selection is to adhere with trigger limitations. To ensure this, we follow the cuts of an ATLAS dijet resonance search~\cite{ATLAS:2019fgd}. First, the two jets are required to have $p_T > 150$~GeV and $|\eta| < 2$. Two more event level cuts are applied. The first is based on half the rapidity separation of the leading jets, $y^* = (y_1 - y_2)/2$. The absolute value of this observable tends to be smaller for $s$-channel processes, such as our resonant signal. To increase signal purity we therefore require $|y^*| < 0.8$. The second requirement is a minimal azimuthal separation between leading jets, $\Delta \phi(j_1, j_2) = |\phi_1 - \phi_2| > 1$, to prevent excessive overlap between the jets. Finally, a lower bound of 1133~GeV on dijet invariant mass $m_{jj}$ is required to ensure compliance with the trigger~\cite{ATLAS:2019fgd}.

\subsubsection{Displaced object selection}
We wish to select for dijet events with displaced objects. The criterion we chose for such events is that at least 20\% of the jet transverse momentum should be carried by tracks that are associated with reconstructed displaced vertices. To suppress contributions from long-lived SM hadrons, vertices with two tracks and vertex mass close to the $\Lambda$ or $K_S^0$ masses (computed with the appropriate particle identity assumptions for the products) are discarded. A summary of event requirements is given in table \ref{ev_req}.

\begin{table}[h]
	\centering
	\setlength{\extrarowheight}{.5ex}
	\begin{tabular}{|c l|}
		\hline
		$p_T^{\rm jet}$ &
		$ > 150 \ \mathrm{GeV}$\\
		
		$|\eta|^{\rm jet}$ &
		$< 2$ \\
		
		$m_{jj}$ &
		$> 1133$ GeV \\
		
		$|y^*|$ &
		$< 0.8$ \\
		
		$\Delta \phi(jj)$ &
		$> 1$ \\
		
		$\sum\limits_{\mathrm{disp. vert.}} p_T^{\mathrm{vertex}} / p_T^{\mathrm{jet}}$ &
		$> 0.2$\\
		\hline
	\end{tabular}
	\caption[Event selection summary]{Event selection summary. Both leading jets (highest $p_T$) must satisfy $p_T$, $\eta$, and displaced vertex requirements.}
	\label{ev_req}
\end{table}

\subsection{Standard Model background}
\label{bm_bkg}

Displaced vertices are primarily a signature of events containing heavy flavor ($b$ or $c$) quarks.  We therefore expect the leading SM background for our analysis to be dijet events where both leading jets are of a heavy flavor. To estimate the background magnitude and composition we generated events in four groups:
\begin{itemize}
\item ``$bb$ events'' contain a pair of $b$-flavored jets at the parton level. These events are primarily $b\bar b$, with smaller contributions from $bb$ and $\bar b\bar b$.
\item ``$cc$ events'' contain a pair of $c$-flavored jets, analogous to the above.
\item ``$bc$ events'' contain one $b$-flavored jet and one $c$-flavored jet.
\item The fourth group consists of the remaining dijet events (which we will call ``other''), which are events that contain only one or no heavy-flavor jets at the parton level.
\end{itemize}
Event generation, including details about detector simulation and vertexing, is described in appendix~\ref{evgen}.

The total background cross section after the selection described in section~\ref{ev_select_section} is $\sim 0.13$~pb.\footnote{To reduce the computational burden, our simulation used leading-order matrix elements, followed by parton showering. However, higher-order QCD corrections, including hard jet radiation, can have some effect on the production cross section, the selection efficiency, and the properties of the two leading jets. While our data-based search methodology is not directly dependent on the simulation details, quantitative claims about the range of scenarios that can be discovered or excluded may be affected.} The leading contribution ($\sim 50\%$) comes from $bb$ events. The next dominant background ($\sim 37\%$) comes from light or semi-light QCD events, i.e., with less than two final state heavy quarks at the parton level (the ``other'' category above). This group is dominated by events with gluons splitting to heavy quarks, a process that is significant in the hard events under consideration~\cite{ATLAS:2012mwf}. The remaining groups, $bc$ and $cc$, account for 10\% and 3\% of the events, respectively. The selection efficiencies of the different groups are summarized in table~\ref{bkg_composition_tbl}.

\begin{table}[h]
	\centering
	\setlength{\extrarowheight}{.5ex}
	\begin{tabular}{|c c c c c c|}
		\hline
		Group &
		$N_{\mathrm{prim}}$ &
		$N_{\mathrm{pass}}$ &
		$\epsilon_{\mathrm{DO}}$ &
		$\sigma_{\mathrm{prim}}$ (pb) &
		$\sigma$ (pb) 
		\\
		\hline\hline 
		$bb$ &
		1066652 &
		100551 &
		0.094 &
		0.71 &
		0.067 
		\\
		\hline 
		$jj$ (``other'') &
		671729 &
		62 &
		$9.2 \cdot 10^{-5}$ &
		530 &
		0.049 
		\\
		\hline 
		$cc$ &
		2059665 &
		27069 &
		0.013 &
		0.98 &
		0.013 
		\\
		\hline 
		$bc$ &
		577405 &
		9163 &
		0.016 &
		0.24 &
		0.0038 
		\\
		\hline
	\end{tabular}
	\caption
	[Selection efficiencies and magnitudes of different SM channels]
	{Selection efficiencies and magnitudes of different SM channels. The cross section after the primary selection, $\sigma_{\mathrm{prim}}$, is derived from the generation level cross section of group events obtained from MadGraph at leading order times the efficiency of the primary selection. Displaced object efficiency is presented with respect to events after primary selection, namely $\epsilon_{\rm DO} = \frac{N_{\mathrm{pass}}}{N_{\mathrm{prim}}}$ (where $N_{\mathrm{prim}}$ and $N_{\mathrm{pass}}$ refer to the numbers of our MC events). The final available cross section is determined according to $\sigma = \epsilon_{\mathrm{DO}} \sigma_{\mathrm{prim}}$.} 
	\label{bkg_composition_tbl}
\end{table}

\subsection{Jet features used for classification}
\label{jettfeats}

The cotraining step of the search requires a choice of jet classification model and jet representation. We chose to represent each jet as a list of high-level features as input to a dense NN model. A complete description of the NN model we used is provided in appendix~\ref{classifier_details}. More complex representations and architectures, such as a long short-term memory (LSTM) network on lists of vertex features, were also considered. In our testing, these were outperformed by the simple dense architecture and therefore abandoned. This could change as the amount of analysis data grows since more data often favors more complex networks.

Jet features include vertex features chosen to represent the properties of displaced objects within a jet and general jet features that encode complementary jet information. We consider the following vertex features: vertex mass, vertex transverse displacement $D_0$ divided by the boost factor $\gamma \beta_T$, fraction of jet's transverse momentum carried by the vertex tracks, and vertex track count. For the features above, in the case of more than one reconstructed vertex, the median value across reconstructed vertices is used. The boost factor, $\gamma \beta_T$, is computed according to
\begin{equation}
	\gamma \beta_T = \frac{p_T^{\mathrm{vertex}}}{m_\mathrm{vertex}}
\end{equation}
where $p_T^{\mathrm{vertex}}$ is the magnitude of the vector sum of $\mathbf{p}_T$s of tracks associated to the vertex. Vertex mass is calculated according to
\begin{equation}
	\label{vert_mass_eq}
	m_{\mathrm{vertex}}^2=\left(\sum_{\mathrm{tracks}}\sqrt{\mathbf{p}_{\mathrm{track}}^2 + m_{\pi^{\pm}}^2}\right)^2 - \left(\sum_{\mathrm{tracks}}\mathbf{p}_{\mathrm{track}}\right)^2
\end{equation}
i.e.\ the tracks are assigned the charged pion mass for estimation of their energy, and the sum is over all tracks associated with the vertex. We also supply the total number of reconstructed vertices in the jet, excluding the primary vertex and the number of particle-flow objects in the jet -- $n_{\mathrm{obj}}$.

In our toy dark sector models that will be described in the next section, the discrimination power of each of these features varies with dark sector parameters. The dark pion mass $m_{\pi'}$ directly affects $m_{\mathrm{vertex}}$. Increasing dark pion mass also decreases the number of vertices per jet and increases the number of tracks per vertex. The dark pion lifetime $\tau_{\pi'}$ directly affects $(D_0/\gamma\beta_T)^{\mathrm{vertex}}$ and also indirectly affects the number of vertices. For larger dark pion lifetime, more displaced vertices are distinguished from the primary vertex and therefore the number of displaced vertices increases.

\section{Benchmark datasets}
\label{benchmark_datasets}

While the search is intended to be largely model independent, it is useful to test the strategy on some examples. Since the detailed physics of confining hidden sectors is not known well and is very model dependent, and the simulation tools are limited too, we consider a set of simplistic toy models, defined as follows.

\subsection{Benchmark hidden sectors}
\label{bm_sig}

We base our toy models on the scenario that is obtained in the \textsc{Pythia8} Hidden Valley module~\cite{Carloni:2010tw} for an $SU(3)$ gauge group with a single quark flavour. We consider fully visible jets, i.e.\ $r_{\mathrm{inv}} = 0$, which manifests as no excessive missing transverse energy. We consider six combinations of the remaining two parameters, with values $(m_{\pi'},  c\tau_{\pi'}) =  \{5~\mathrm{GeV}, 10~\mathrm{GeV}\} \times \{0.1~\mathrm{mm}, 0.2~\mathrm{mm}, 0.3~\mathrm{mm}\}$. Other mass parameters of the dark sector---confinement scale, constituent dark quark mass, and vector meson mass---were scaled with $m_{\pi'}$, starting at $\Lambda_{\mathrm{QCD}'}=5~\mathrm{GeV},\ m_{q'}=5~\mathrm{GeV}$, and $m_{\rho'}=10.5~\mathrm{GeV}$ for $m_{\pi'}=5~\mathrm{GeV}$. The probability for creating dark vector mesons is kept at its default value of $0.75$. We assume the dark vector mesons decay promptly and exclusively to dark pion pairs: $\rho' \to \pi'\pi'$. Our simulated dark pions decay exclusively to SM down quark-antiquark pairs: $\pi' \to d\bar d$. Decays to heavier flavor quarks are in principle possible for $m_{\pi'}$ values considered and perhaps even motivated by helicity suppression. However, we found such scenarios less interesting as they produce many additional displaced vertices from the decays of the heavy flavor quarks, amounting to a signal too distinct from QCD background. The benchmark Hidden Valley parameters are summarized in table~\ref{signal_params}.

\begin{table}[h]
	\centering
	\setlength{\extrarowheight}{.5ex}
	\begin{tabular}{|c c |}
		\hline
		Gauge group &  SU(3) \\
		$\Lambda_{\mathrm{QCD}'}$ & 5 / 10 GeV\\
		$n_{q'}$ & 1 \\
		$m_{q'}$ & 5 / 10 GeV\\
		$m_{\pi'}$ & 5 / 10 GeV \\
		$m_{\rho'}$& 10.5 / 21 GeV \\
		$c\tau_{\pi'}$ & 0.1 / 0.2 / 0.3 mm \\
		$r_{\mathrm{inv}}$ & 0\\
		\hline
	\end{tabular}
	\caption{Hidden Valley parameters used for the six benchmark signal configurations.}
	\label{signal_params}
\end{table}

We test the sensitivity of our proposed search to resonant dijet events produced via $pp \to Z' \rightarrow q' \bar{q}'$ with $m_{Z'}=2$~TeV. We assume an interaction Lagrangian of the following form:
\begin{equation}
	{\cal L} \supset -g_q Z'_\mu \bar q\gamma^\mu q - g_{q'} Z'_\mu \bar q'\gamma^\mu q' \,,
\end{equation}
where the first term describes the $Z'$ coupling to SM quarks and the second to dark quarks.

\subsection{Benchmark datasets}
\label{bm_dataset}

Our background dataset contains $106k$ events that passed the full event selection in the dijet invariant mass range $1400~\mbox{GeV} \leq m_{jj} \leq 2400~\mbox{GeV}$. The majority, $89k$, are $bb$ events and the rest, $17k$, are $cc$ events. As can be seen in table~\ref{bkg_composition_tbl}, these two channels combined account for the majority of QCD events that pass selection. We would ideally simulate the entire QCD dijet sample (rather than only $bb$ and $cc$); however this is too computationally expensive for us due to low selection efficiencies of the displaced objects cut. If we rescale the cross section so that the total background cross section is correct, then based on the analysis of section~\ref{bm_bkg} this example corresponds to an integrated luminosity of $\sim800 \ \mathrm{fb}^{-1}$ available for the analysis.

We will analyze in detail the example of a $Z'$ with mass $m_{Z'}=2$~TeV and a negligible width relative to the experimental dijet invariant mass resolution. Motivated by the shape of the resulting dijet invariant mass distribution (see, e.g., figure~\ref{mjj:mjj_all}), whose width is not very model dependent since it is dominated by the experimental resolution, we define the signal region to be the invariant mass range $m_{jj} \in [1600, 2000]$~GeV.\footnote{This choice of the mass window is based on the region in which the resonant contribution appears in our \textsc{Delphes} simulation. In an actual experimental analysis, this choice should be reconsidered based on a more accurate simulation of the corresponding detector and jet energy scale corrections that are applied to the reconstructed jets.} We define the sidebands as $m_{jj} \in [1400, 1600) \cup (2000, 2400]$~GeV. These boundaries are chosen such that the sidebands and the signal region contain comparable numbers of background events. Approximately 20\% of the background events are in the signal region. Signals, one of the six hidden sector configurations described in section~\ref{bm_sig}, are injected to this background. Signal size, which we will vary, will be presented in terms of signal fraction $f_S = N_S/ (N_B + N_S)$, where $N_B$ and $N_S$ and the background and signal event count in the entire mass range (signal region and sidebands) after event selection. 

The feature distributions for the different benchmark signals (and the background) are provided in appendix~\ref{feat_hists}. The most discriminating features for this set of benchmarks are the number of objects in the jet, vertex count, and transverse momentum fraction. One can also see that as dark pion displacement increases, from 0.1~mm to 0.3~mm, the number of signal vertices increases because more dark pions decay outside the primary vertex resolution. Vertex mass is a stronger discriminator for the 10~GeV dark pion mass in comparison to the 5~GeV case.

\section{Example search}
\label{example_search}

In this section we present results of an example search conducted on a simulated benchmark dataset. We provide a detailed account for the case of a dark sector with $m_{\pi'} = 10$~GeV and $c\tau_{\pi'} = 0.2$~mm with a signal fraction $f_S = 0.5\%$, where the number of signal events is $N_S =  N_B \cdot \frac{f_S}{1-f_S} = 530$~events. We provide aggregated results for different signal fractions of all other benchmark signals.

\subsection{Weak jet classifier}
\label{weak}

We used the number of particle-flow objects, $n_{\mathrm{obj}}$, of each jet to assign a signal-rich or background-rich (weak) label for the other jet, as described in section~\ref{search_strategy}. Background-like threshold, $n^B_{\mathrm{obj}}$, was taken to be the $25\%$ (lower) quantile for the number of particle-flow objects. Signal-like threshold, $n^S_{\mathrm{obj}}$, was taken to be the $75\%$ (upper) quantile for the number of particle-flow objects (amounting to $n^B_{\mathrm{obj}}$ = $n^S_{\mathrm{obj}}$). The softer signal-like threshold is complemented by the invariant mass region selection so that after both cuts the signal- and background-rich labels are approximately balanced. The thresholds were chosen after trying a number of alternatives and finding that results are not very sensitive to this choice. Using tighter signal- and background-rich thresholds amounts to a higher effective signal fraction for training. This comes at the cost of less data available for training. A quantification of this tradeoff is left for future works.

Jet constituent count thresholds corresponding to the chosen quantiles are $n^{\rm thresh}_{\mathrm{obj}} = 24$ for cuts on $j_1$ and $n^{\rm thresh}_{\mathrm{obj}} = 25$ for cuts on $j_2$. The difference stems from a slightly higher object multiplicity for second jets. These values were unaffected by the small signal fractions considered and are therefore the same for all signals and all signal fractions. From cutting on $j_1$ and $j_2$ constituent counts, and requiring that signal-rich jets come from signal-region events, we obtain two background-rich and two signal-rich samples.

In the 0.5\% signal fraction case of the $(m_{\pi'},  c\tau_{\pi'}) = (10~\mathrm{GeV},\ 0.2~\mathrm{mm})$ signal, these cuts leave 26292 (23906) background-rich jets and 27824 (28432) signal-rich jets from cuts on $j_2$ ($j_1$) constituent counts. From an initial 0.5\% signal fraction in the entire dataset, the enriched signal fractions are 1.59\% (1.56\%) in the signal-rich samples and 0\% (0\%) in the background-rich samples from cuts on $j_2$ ($j_1$) multiplicities.

\subsection{Weakly supervised event classifier}
\label{weaksup}

\begin{figure}[t]
	\centering
	\begin{subfigure}[b]{0.32\textwidth}
		\centering
		\includegraphics[width=\textwidth]{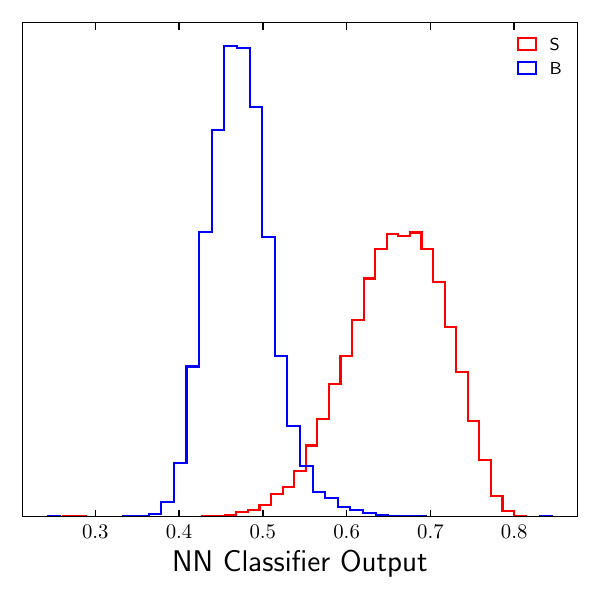}
		\caption{$j_1$ classifier}
		\label{nnout:j1}
	\end{subfigure}
	\begin{subfigure}[b]{0.32\textwidth}
		\centering
		\includegraphics[width=\textwidth]{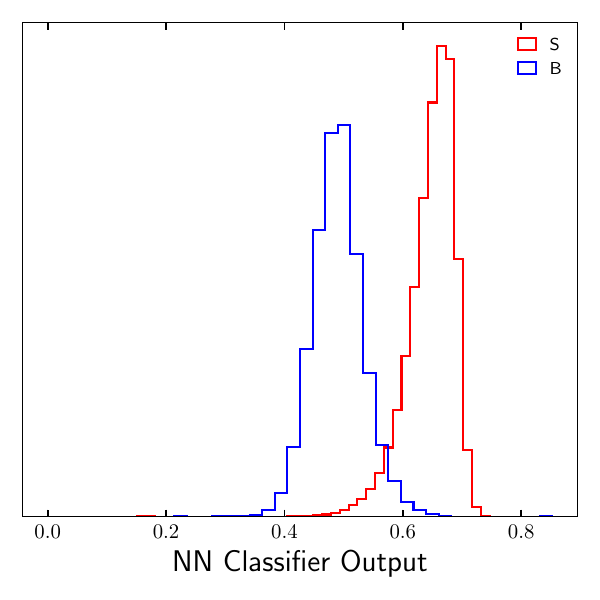}
		\caption{$j_2$ classifier}
		\label{nnout:j2}
	\end{subfigure}
	\begin{subfigure}[b]{0.32\textwidth}
		\centering
		\includegraphics[width=\textwidth]{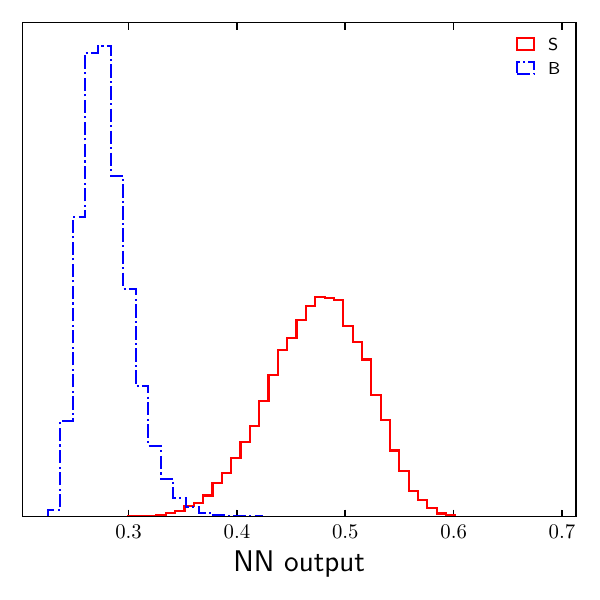}
		\caption{combined: $j^{\rm pred}_1 \cdot j^{\rm pred}_2$}
		\label{nnout:j3}
	\end{subfigure}
	\caption[Distribution of NN output]
	{NN output distributions for $j_1$, $j_2$, and combined classifiers, for the scenario with $(m_{\pi'},  c\tau_{\pi'}) = (10~\mathrm{GeV},\ 0.2~\mathrm{mm})$, $f_S = 0.5\%$.}
	\label{NNout_semi}
\end{figure}

\begin{figure}[ht]
	\centering
	\includegraphics{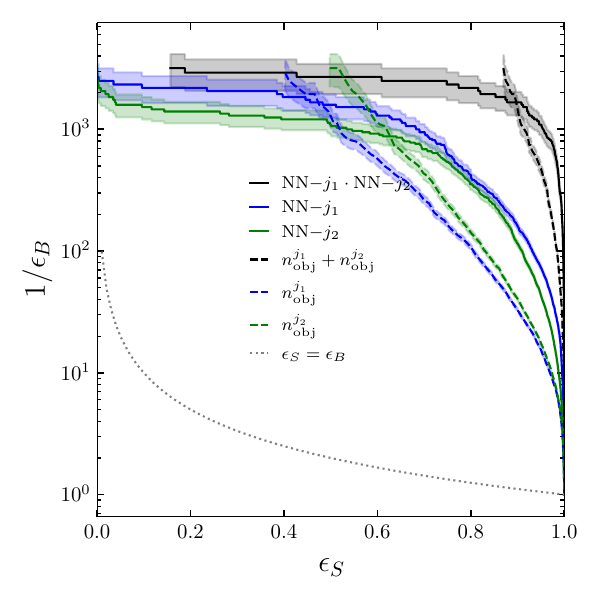}
	\caption[Event and jet level ROCs for $(m_{\pi'},  c\tau_{\pi'}) = (10~\mathrm{GeV},\ 0.2~\mathrm{mm})$ signal at $f_S=0.5\%$]{Solid curves are ROCs for NN jet classifiers trained using cotraining, and the event classifier obtained from their product, for the scenario with $(m_{\pi'},  c\tau_{\pi'}) = (10~\mathrm{GeV},\ 0.2~\mathrm{mm})$, $f_S = 0.5\%$. Dashed curves are ROCs for the weak jet classifiers (constituent count of each jet) and the event classifier obtained from their sum. Shaded areas signify $1\sigma$ statistical uncertainty, where for a cut leaving $N_B$ background events out of a total $N_B^0$ background events, we used $\sigma = N_B^0/N_B^2 \cdot \sqrt{N_B}$. The curves were terminated at $N_B<10$ $(1/\epsilon_B=3500)$.}
	\label{ROC_semi}
\end{figure}

\begin{figure}[ht]
	\centering
	\includegraphics[scale=1]{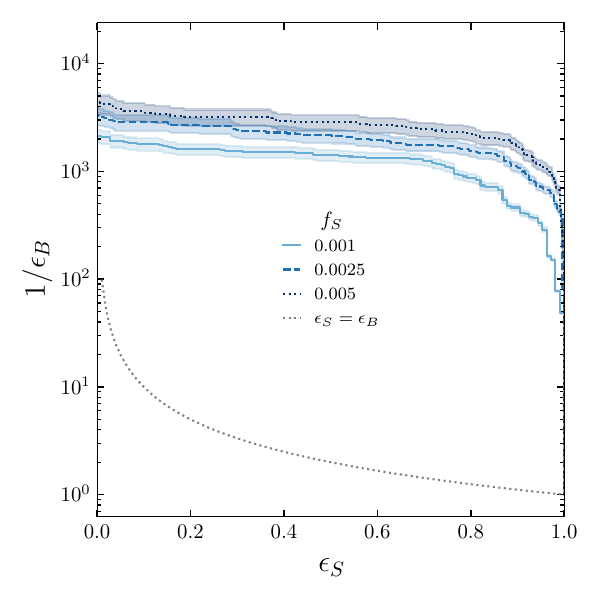}
	\caption[Event classifier ROCs for $(m_{\pi'},  c\tau_{\pi'}) = (10~\mathrm{GeV},\ 0.2~\mathrm{mm})$ signal at $f_S=0.1 - 0.5\%$]{ROCs comparing classifiers trained using weakly supervised learning with varying signal fractions $f_S$ of the benchmark signal with $(m_{\pi'},  c\tau_{\pi'}) = (10~\mathrm{GeV},\ 0.2~\mathrm{mm})$.}
	\label{sf_compare}
\end{figure}

\begin{figure}[ht]
	\centering
	\includegraphics[scale=1]{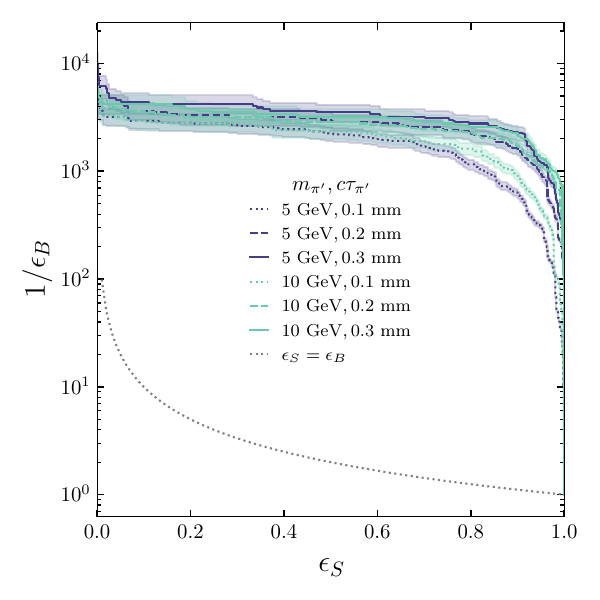}
	\caption[Event classifier ROCs for all benchmark signals at $f_S=0.5\%$]{ROCs comparing classifier discrimination for different benchmark signals. All classifiers were trained with $f_S = 0.5\%$.}
	\label{sigtype_compare}
\end{figure}

After applying the weak cuts to the $0.5\%$ signal fraction case of the signal with $(m_{\pi'},  c\tau_{\pi'}) = (10~\mathrm{GeV},\ 0.2~\mathrm{mm})$, $51\%$ ($49\%$) of jets were assigned weak labels according to $n^{j_1}_{\rm obj}$ ($n^{j_2}_{\rm obj}$). Of these events, $10\%$ are put aside for validation to avoid overfitting. A classifier, described in appendix~\ref{classifier_details}, is trained to distinguish between the remaining $46\%$ ($44\%$) of events using $j_1$ ($j_2$) features and the weak labels. After weak-label assignment and putting aside of the validation set, $48704$ jets and $47104$ jets are available for training $j_1$ and $j_2$ classifiers, respectively. The classifiers were trained for 100 epochs. Learning curves are presented in figure~\ref{learncurve:both}. 
To evaluate the classifiers' performance, a new dataset of $35k$ signal and $35k$ background events was generated. NN outputs and receiver operating characteristic (ROC) curves for the $0.5\%$ signal fraction case of the $(m_{\pi'},  c\tau_{\pi'}) = (10~\mathrm{GeV},\ 0.2~\mathrm{mm})$ signal are shown in figures~\ref{NNout_semi} and~\ref{ROC_semi}. ROCs comparing discrimination of classifiers trained on different signal fractions are shown in figure~\ref{sf_compare}. As expected, classifier performance deteriorates as $f_S$ is decreased. Still, even at $f_S=0.1\%$, the classifier is very powerful. However, going to much lower signal fractions is not relevant because they will not be detectable eventually in the bump hunt procedure that is discussed in the next subsection. ROCs comparing the outcomes for the different benchmark signals with 0.5\% signal fraction are shown in figure~\ref{sigtype_compare}.

\subsection{Identifying and quantifying an excess}

Our null hypothesis, which we will confirm in the following, is that the dijet invariant mass distribution of the background after the NN cut is still well described by a smoothly decreasing function. We construct the following test statistic to probe for deviations from this hypothesis due to a possible signal. We bin the events (with a bin size of 50~GeV in our example) and fit the sidebands to the following three-parameter function 
\begin{equation}
	\label{fit_func}
	\frac{dN}{dm_{jj}} = p_0\frac{(1-m_{jj}/\sqrt{s})^{p_1}}{(m_{jj}/\sqrt{s})^{p_2}},
\end{equation}
also used in ATLAS~\cite{ATLAS:2017zuf} and CMS~\cite{CMS:2016rqm}. The fit parameters $p_i$ were constrained to positive values. We estimate the number of expected events in the signal region using the fit and compare it to the measured number of events in the signal region. Our test statistic is the excess
\begin{equation}
	\label{tstat}
	t = 
	\frac{
		N^{\mathrm{sig. reg.}}_{\mathrm{meas}} - N^{\mathrm{sig. reg.}}_{\mathrm{exp}}}
	{\sqrt{\sigma^2_{\mathrm{meas}} + \sigma^2_{\mathrm{exp}}}} \;.
\end{equation}
The uncertainty in the measured counts is estimated by Poisson statistics as $\sigma^2_{\mathrm{meas}} = N^{\mathrm{sig. reg.}}_{\mathrm{exp}}$. The uncertainty in the expected counts is obtained by linearly propagating parameter fit uncertainties. Further details of this procedure are provided in appendix~\ref{fit_procedure}. We obtain test statistic values for different cut efficiencies of the NN. To avoid training many classifiers, as would be done in the $k$-fold procedure described in section~\ref{search_strategy}, we use the entire $106k$ event dataset for semisupervised training and continue with inference on a new (same size) dataset. This is similar to the $k$-fold procedure for a large enough $k$.

\begin{figure}[ht]
	\centering
	\vspace{-1\baselineskip}
	\begin{subfigure}[b]{0.47\textwidth}
		\centering
		\includegraphics[width=\textwidth, page=1,trim={0 0.8cm 0 0.7cm},clip]{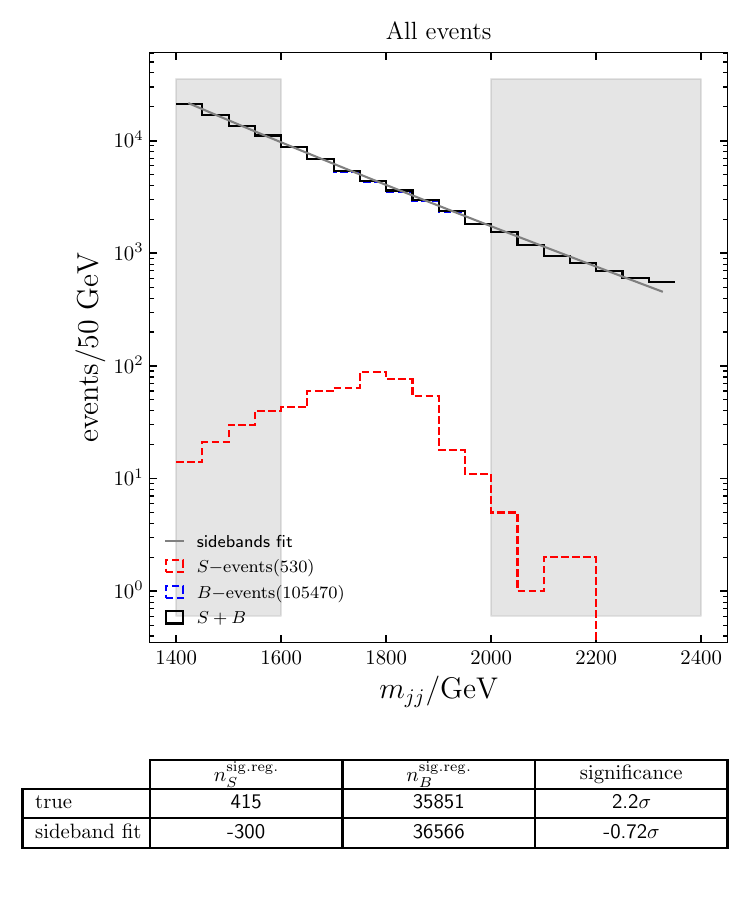}
		\caption{all events}
		\label{mjj:mjj_all}
	\end{subfigure}
	\begin{subfigure}[b]{0.47\textwidth}
		\centering
		\includegraphics[width=\textwidth, page=17,trim={0 0.8cm 0 0.87cm},clip]{bumphunt.pdf}
		\caption{$\epsilon_D=2\%$}
		\label{mjj:mjj_2}
	\end{subfigure}
	\par\bigskip
	\begin{subfigure}[b]{0.47\textwidth}
		\centering
		\includegraphics[width=\textwidth, page=16,trim={0 0.8cm 0 0.87cm},clip]{bumphunt.pdf}
		\caption{$\epsilon_D=1\%$}
		\label{mjj:mjj_1}
	\end{subfigure}
	\begin{subfigure}[b]{0.47\textwidth}
		\centering
		\includegraphics[width=\textwidth, page=14,trim={0 0.8cm 0 0.87cm},clip]{bumphunt.pdf}
		\caption{$\epsilon_D=0.6\%$}
		\label{mjj:mjj_0.6}
	\end{subfigure}
	\caption[Invariant mass distribution after various cuts]
	{Invariant mass spectrum of events passing NN cut with varying selection efficiency $\epsilon_D$. The sidebands are shaded. The significance in the ``true'' rows corresponds to true $n_S/\sqrt{n_B}$ within the signal region. The significance in the ``sideband fit'' rows is obtained from the fit parameters according to eq.~\eqref{tstat}.}
	\label{mjj:mjj}
\end{figure}

Let us now exemplify the search with one realization of a background and signal sample. Invariant mass distributions subject to NN cuts of varying efficiency are presented in figure~\ref{mjj:mjj} for the $(m_{\pi'},  c\tau_{\pi'}) = (10~\mathrm{GeV},\ 0.2~\mathrm{mm})$ signal with $f_S = 0.5\%$. The invariant mass spectrum of the entire dataset after the event selection described in section~\ref{ev_select_section} is shown in figure~\ref{mjj:mjj_all}. The test statistic significance prior to any further cut is $-0.72 \sigma$. (The negative sign indicates a downward deviation.) The invariant mass spectra after applying the NN cuts with $\epsilon_D=2\%$, $1\%$, $0.6\%$ are shown in figures~\ref{mjj:mjj_2},~\ref{mjj:mjj_1},~\ref{mjj:mjj_0.6}, respectively. Apart from the significances obtained with fit to the sidebands, the information tables at the bottom of these plots show also the significance estimates that would be obtained from the naive calculation of $n_S/\sqrt{n_B}$ in the signal region. As expected, since it does not account for the statistical fluctuations in the sidebands and for the potential contributions in the sidebands due to the signal tails, the naive estimate of the significance is unrealistic, and it is crucial to simulate the fit to the sidebands like we did. 

\begin{figure}[ht]
	\centering
	\includegraphics{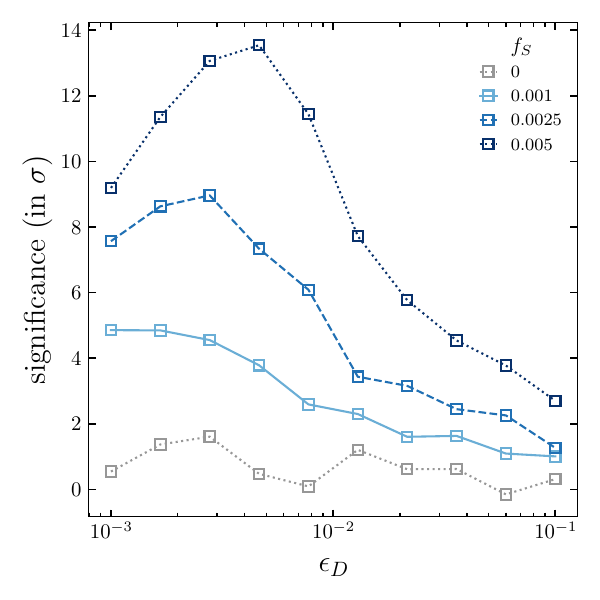}
	\caption[Bump significance vs. selection efficiency for four signal fractions]{Bump significance as a function of selection efficiency at four signal fractions for $(m_{\pi'},  c\tau_{\pi'}) = (10~\mathrm{GeV},\ 0.2~\mathrm{mm})$ signal.}
	\label{sf_compare_signif}
	\centering
	\includegraphics{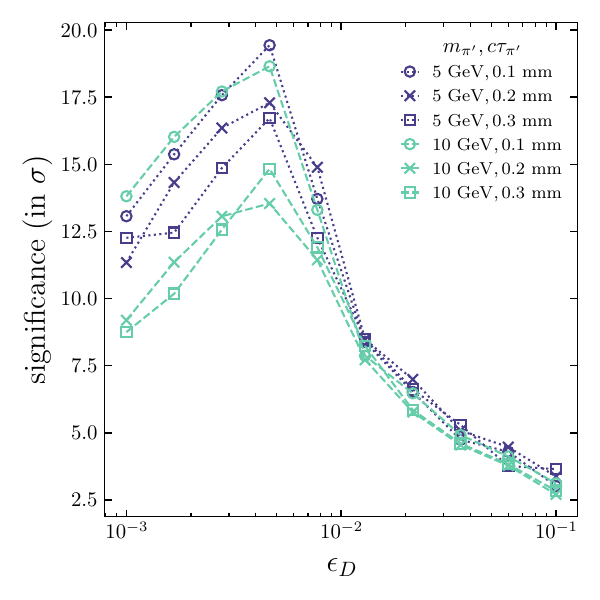}
	\caption[Bump significance vs. selection efficiency for six benchmark signals]{Bump significance as a function of selection efficiency for the six benchmark signals at $f_S = 0.5\%$.}
	\label{sigtype_compare_signif}
\end{figure}

Significance as a function of $\epsilon_D$ for different signal fractions is presented in figure~\ref{sf_compare_signif}. The significance peaks at $\epsilon_D \sim f_S$, peaking at higher data efficiencies for greater signal fractions. For $f_S=0.25\%$ and $f_S=0.5\%$ the signals pass the discovery threshold of $5\sigma$ whereas the $f_S=0.1\%$ signal falls short. Significance obtained for different benchmark signals with $f_S = 0.5\%$ is presented in figure~\ref{sigtype_compare_signif}. All benchmark signals are discoverable at this signal fraction.

We also test for the bump significance in a dataset with no signal. This corresponds to the $f_S=0$ curve in figure~\ref{sf_compare_signif}. The significance fluctuates between $\sim 0$ and $2$ $\sigma$ for the values of $\epsilon_D$ considered, which reassures us that no large, spurious bump is carved in the analysis. However, a similar significance trend was observed in a second background realization we tested, suggesting that some $\sim +1\sigma$ bias exists in our significance estimation. A more detailed study, which would involve generating a large number of background realizations, will be needed to quantify the size of this apparent bias more precisely. Additionally, one could explore whether the bias could be reduced by using a different fitting function. More sophisticated methods to reduce sculpting (e.g., along the lines of ref.~\cite{Benkendorfer:2020gek}) could also be explored.

While the weakly supervised machine learning method outlined here is working well, it is interesting to ask how it performs relative to simpler methods. An obvious comparison in the case of our benchmark models is to cutting on the jet constituent multiplicity variable. Instead of the loose cut on the multiplicity that was used in the ML approach for producing the weak labels, we apply tight cuts on the sum of the multiplicities of the two jets and use the same sidebands fit procedure. Table~\ref{results_table} summarises the bump significance obtained when cutting on the output of the weakly supervised event classifier and when cutting on the sum of object multiplicities of both jets, for all benchmark signals. The $(m_{\pi'},  c\tau_{\pi'}) = (5~\mathrm{GeV},\ 0.3~\mathrm{mm})$ signal was discovered at $f_S = 0.25\%$ and all the rest were discovered at $f_S = 0.1\%$. For these signal fractions it was usually the case that cutting on multiplicity slightly outperformed the NN. (Note, however, that significance values like $5\sigma$ and higher are somewhat uncertain because they assume the fluctuations due to the fit uncertainties to remain Gaussian far on the tails. Also, the bias discussed in the previous paragraph needs to be quantified for both methods. Therefore, small differences should not be taken too seriously.) The exception is the $(m_{\pi'},  c\tau_{\pi'}) = (5~\mathrm{GeV},\ 0.3~\mathrm{mm})$ signal that was discovered with higher significance by the NN at $f_S=0.25\%$.

\clearpage

\begin{table}[h]
	\centering
	\setlength{\extrarowheight}{.5ex}
	\begin{tabular}{ccccc}
		\toprule
		$f_S$ &    $m_{\pi'}$ &   $c\tau_{\pi'}$ & $\max{\sigma_{\rm NN}}$ & $\max{\sigma_{n_{\rm obj}}}$ \\
		\midrule
		0.1\% &  5~GeV & 0.1~mm &  5.4~$\sigma$ &    7.2~$\sigma$ \\
		0.1\% &  5~GeV & 0.2~mm &  5.2~$\sigma$ &    5.9~$\sigma$ \\
		0.1\% &  5~GeV & 0.3~mm &  4.3~$\sigma$ &    4.8~$\sigma$ \\
		0.25\% &  5~GeV & 0.3~mm &  9.6~$\sigma$ &    8.6~$\sigma$ \\
		0.1\% & 10~GeV & 0.1~mm &  5.4~$\sigma$ &    6.8~$\sigma$ \\
		0.1\% & 10~GeV & 0.2~mm &  4.9~$\sigma$ &    7.2~$\sigma$ \\
		0.1\% & 10~GeV & 0.3~mm &    4~$\sigma$ &    5.6~$\sigma$ \\
		\bottomrule
	\end{tabular}
	\caption[Estimated discovery reach for all benchmark signals]{Comparison between object multiplicity cut and weakly supervised NN cut. The columns $\max{\sigma_{\rm NN}}$ and $\max{\sigma_{n_{\rm obj}}}$ correspond with the maximum bump significance across selection efficiencies $\epsilon_D$ for cuts on NN output and cuts on the sum of object multiplicities of both jets, respectively. For this we consider ten values of $\epsilon_D$ spaced log uniformly in the range $[0.001, 0.1]$.} 
	\label{results_table}
\end{table}

\section{Summary and conclusions}
\label{conclusions}

A hidden (``dark'') confining sector may reveal itself at the LHC in the form of anomalous jets, dubbed dark jets, whose properties are very model dependent. In this work we considered dark sectors with dark hadron lifetimes similar to heavy-flavor QCD quarks. A main feature of jets arising from such a sector is displaced vertices from the decays of dark hadrons. We propose using the features of reconstructed vertices to further capture the properties of the displaced objects. The dark sector scenarios we consider are complementary to the ones considered in most of the papers on the subject, which assume the presence of missing energy or very large vertex displacements or do not take advantage of displaced vertices.

The wealth of data collected at the LHC offers an opportunity to harness machine learning to discriminate BSM from SM signatures. A traditional approach to doing so is using MC simulations of signal (or a mix of signals) events and of SM events to train a NN. This paradigm has drawbacks. There are large uncertainties in simulating events, introduced by modeling uncertainties of nonperturbative QCD processes (and in our case also those of the dark confining sector) as well as detector modeling. Another drawback is a lack of generality which translates to reduced sensitivity (if any) to signals not used for training. This is a problem when sensitivity to a wide range of signals is required. Dark sector details are largely unconstrained, allowing for a wide range of dark jet signatures. In this work we propose using the weakly supervised method Tag N' Train in searches for dark jets with displaced vertices. Tag N' Train is a weakly supervised method to obtain a dijet event classifier. The procedure starts with a weak jet classifier. We propose using a cut on jet constituent multiplicity for this stage. This choice makes use of the fact that many dark sector models produce high multiplicity jets. Using the weak labels obtained from the weak classifier, two classifiers are trained, one for each of the two leading jets in the event. We use dense NN supplied with displaced vertex features, including number of displaced vertices, vertex transverse displacement, vertex mass, number of associated tracks, and the fraction of transverse momentum carried by the vertex out of total jet transverse momentum. Jet constituent multiplicity was also supplied.

We tested this procedure on simulated events with a set of toy dark sector scenarios. We showed that the vertex features can be good discriminators between heavy flavor quark jets and dark jets. We demonstrated a concrete example of a search for resonant dark jet pairs with displaced vertices. The search is conducted in the form of a bump hunt where different mass hypotheses are tested separately. We presented a detailed analysis of the example of a 2~TeV resonance. The resonance mass hypothesis was incorporated in the weak classifier---only jets coming from events within the signal region in invariant mass were candidates to be assigned the signal-rich label. After training the NNs and applying them to simulated data, the significance of the bump was estimated for different NN selection efficiencies. The semisupervised classifier succeeded in learning from auxiliary features specific to the signal that was present in the data for signal fractions as small as $0.1\%$.

However, at least for the range of examples we examined, the sensitivity of our machine learning method turned out to be comparable (with the details of the comparison depending on the model) to what can be achieved by using the number of objects in the two jets, which by itself is a search that has never been done and is worth pursuing. One cause of the NN not offering a big advantage is the low signal fractions. The discrimination power of CWoLa often deteriorates with decreasing signal fraction while a cut on the number of objects in the jets is unaffected. The effective signal fraction can always be increased by tightening the thresholds of the weak classifier. However, this comes at the cost of fewer events available for training the NN. Therefore, this method might improve as more data is collected and available for analysis. 

It could be interesting to extend this method to dark sectors with promptly decaying dark hadrons, where a very different set of features and different backgrounds will be relevant. Another interesting direction would be to consider nonresonant dijet production, where the Tag N' Train method naturally remains applicable.

\acknowledgments

We have greatly benefited from numerous conversations with Hugues Beauchesne, whose insights have contributed significantly to this work. This research was supported in part by the Israel Science Foundation (grants no.~780/17 and 1666/22) and the United States~- Israel Binational Science Foundation (grant no.~2018257). The work of D.B.\ was also supported by the Kreitman Postdoctoral Fellowship and National Postdoctoral Fellowship (NPDF), SERB, PDF/2021/002206, Government of India.

\appendix

\section{Event generation}
\label{evgen}

Parton level events at collider energy of 13~TeV were generated using \textsc{MadGraph5}~\cite{Alwall:2014hca} with the NN23LO1~\cite{Ball:2013hta} PDF set. A massive $Z'$ mediator with couplings to SM quarks and dark quarks was implemented using the Universal FeynRules Output (UFO) files from~\cite{Cohen:2017pzm}. Some parton-level cuts, softer than the eventual selection cuts of section~\ref{ev_select_section}, were applied in MadGraph to save computation time for the background: jet $p_T>100$~GeV, $m_{jj}>1$~TeV, jet $|\eta|<3$, and $\Delta R(j_1, j_2)>1$. Showering and hadronization were simulated using \textsc{Pythia8}~\cite{Sjostrand:2014zea}. Dark-sector showering was done using \textsc{Pythia8}'s Hidden Valley module~\cite{Carloni:2010tw}. Detector simulation was conducted with \textsc{Delphes}~3~\cite{deFavereau:2013fsa} using the ATLAS detector card with added track smearing according to~\cite{CMS:2014pgm}. Jets were reconstructed from calorimeter deposits using the anti-$k_T$ algorithm~\cite{Cacciari:2008gp} with jet radius $R=0.7$. Particle-Flow\footnote{Particle Flow is an algorithm to reconstruct track and calorimeter tower measurements into a list of electrons, muons, charged hadrons, neutral hadrons, and photons.} constituents were then assigned to jets based on their angular distance ($\Delta R$) from the axes of the reconstructed jets. Vertices were reconstructed with the Adaptive Vertex Fitting algorithm (AVR)~\cite{Fruhwirth:2007hz} using default parameters ($\sigma_{\rm cut, p}=2$, $\sigma_{\rm cut, s}=6$, and  $w_{\min}=0.5$), implemented in the RAVE toolkit~\cite{5734880}. All event tracks were used for primary vertex reconstruction while only tracks belonging to a given jet were used to find secondary vertices.

\section{Neural network architecture}
\label{classifier_details}

\begin{figure}[ht]
	\centering
	\begin{subfigure}[b]{0.49\textwidth}
		\centering
		\includegraphics[width=\textwidth]{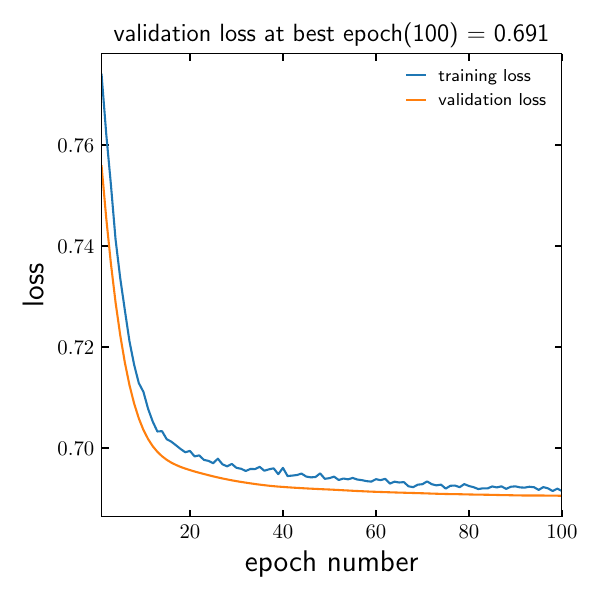}
		\caption{$j_1$ classifier}
		\label{learncurve:j1}
	\end{subfigure}
	\begin{subfigure}[b]{0.49\textwidth}
		\centering
		\includegraphics[width=\textwidth]{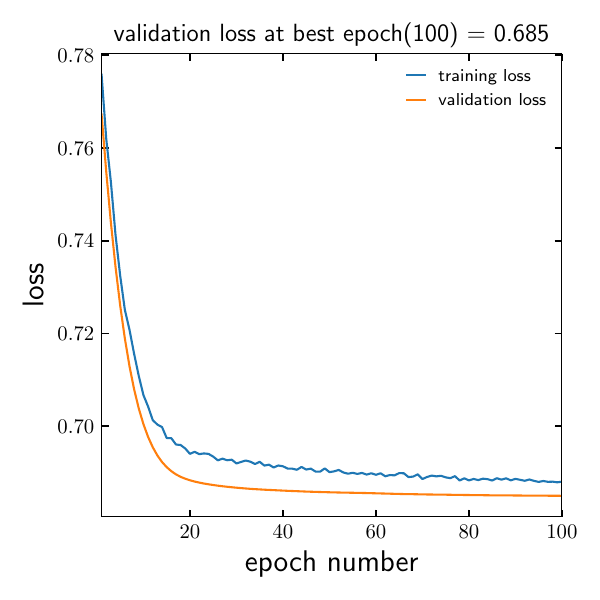}
		\caption{$j_2$ classifier}
		\label{learncurve:j2}
	\end{subfigure}
	\caption[Learning curves for weakly-supervised jet classifiers]
	{Learning curve of $j_1$ (left) and $j_2$ (right) classifiers training on $(m_{\pi'},  c\tau_{\pi'}) = (10~\mathrm{GeV},\ 0.2~\mathrm{mm})$ signal with $f_S = 0.5\%$. Since the validation set is evaluated without use of the dropout layers it is not surprising that the validation set loss is smaller than the training set loss.}
	\label{learncurve:both}
\end{figure}

We use a dense neural network architecture built and trained using Keras~\cite{chollet2015keras} with TensorFlow~\cite{tensorflow2015-whitepaper} backend. The network has four hidden layers with 32, 16, 16, 4 nodes, respectively. These parameters were coarsely optimized to avoid over/under fitting.  The first hidden layer activation is the Leaky Rectified Linear Unit (LeakyReLu). For the remaining three layers Exponential Linear Units (ELU) were used. A sigmoid function was applied to the output. Each hidden layer except the last was followed by a dropout layer with a rate of 0.1. A network summary is provided in table~\ref{nn_summary}. Binary cross-entropy loss function and Adam optimizer were used for training. Each input feature was globally shifted and scaled according to the sample mean and standard deviation of the training data. These scale and shift values are saved. When new data is to be inferred by the classifier it is scaled and shifted by the same values. Examples of learning curves are shown in figure~\ref{learncurve:both}.

\begin{table}[h]
\centering
\setlength{\extrarowheight}{.5ex}
\begin{tabular}{lcr}
\toprule
Layer (type)             & Output shape     &       \# Parameters  \\
\midrule
Layer-1 (Dense)          & (None, 32)       &        224       \\
activation (LeakyReLU)   & (None, 32)       &        0         \\
dropout (Dropout)        & (None, 32)       &        0         \\
Layer-2 (Dense)          & (None, 16)       &        528       \\
activation (ELU)         & (None, 16)       &        0         \\
dropout (Dropout)        & (None, 16)       &        0         \\
Layer-3 (Dense)          & (None, 16)       &        272       \\
activation (ELU)         & (None, 16)       &        0         \\
dropout (Dropout)        & (None, 16)       &        0         \\
Layer-4 (Dense)          & (None, 4)        &        68        \\
activation (ELU)         & (None, 4)        &        0         \\
Output (Dense)           & (None, 1)        &        5         \\
\hline
\multicolumn{3}{c}{Total parameters: 1,097} \\
\bottomrule
\end{tabular}
\caption[NN summary]{NN summary.}
\label{nn_summary}
\end{table}

\FloatBarrier

\clearpage
\section{Feature distributions}
\label{feat_hists}

This appendix presents the feature distributions for the benchmark signals and the background based on events in the mass region $m_{jj} \in [1400, 2400]$~GeV after event selection.

\subsection{\texorpdfstring{
		$(m_{\pi'},  c\tau_{\pi'}) = (5~\mathrm{GeV},\ 0.1~\mathrm{mm})$
	}
	{
		dark pion mass 5 GeV, dark pion lifetime 0.1 mm
	}
}
\subplotfeats{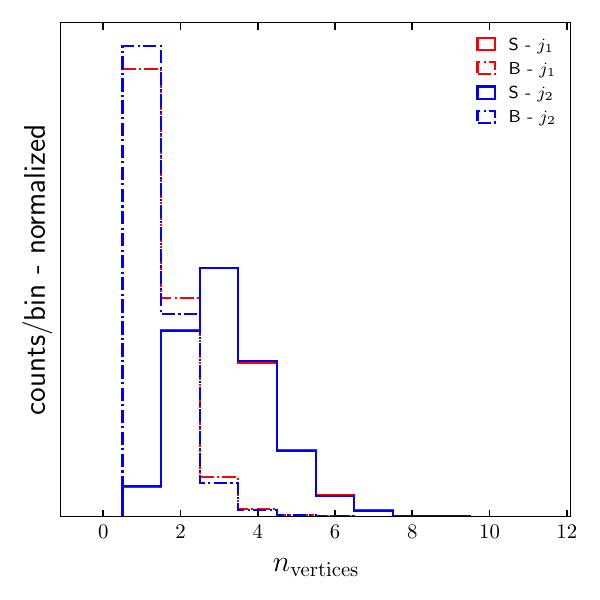}{$(m_{\pi'},  c\tau_{\pi'}) = (5~\mathrm{GeV},\ 0.1~\mathrm{mm})$}
\FloatBarrier
\clearpage

\subsection{\texorpdfstring{
		$(m_{\pi'},  c\tau_{\pi'}) = (5~\mathrm{GeV},\ 0.2~\mathrm{mm})$
	}
	{
		dark pion mass 5 GeV, dark pion lifetime 0.2 mm
	}
}
\subplotfeats{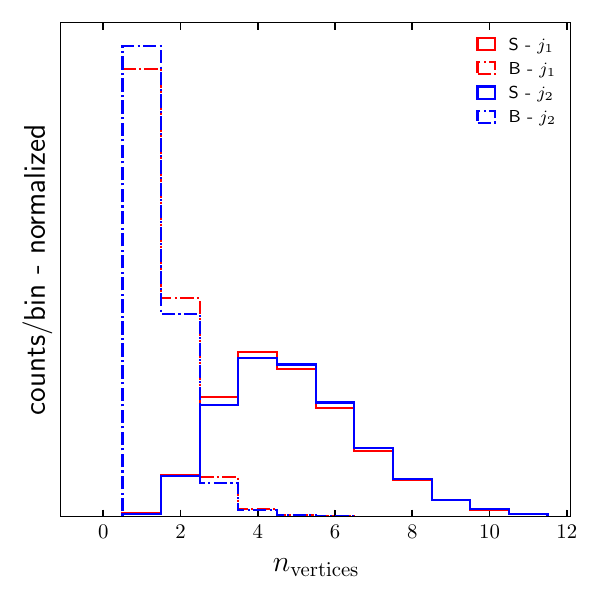}{$(m_{\pi'},  c\tau_{\pi'}) = (5~\mathrm{GeV},\ 0.2~\mathrm{mm})$}
\FloatBarrier
\clearpage

\subsection{\texorpdfstring{
		$(m_{\pi'},  c\tau_{\pi'}) = (5~\mathrm{GeV},\ 0.3~\mathrm{mm})$
	}
	{
		dark pion mass 5 GeV, dark pion lifetime 0.3 mm
	}
}
\subplotfeats{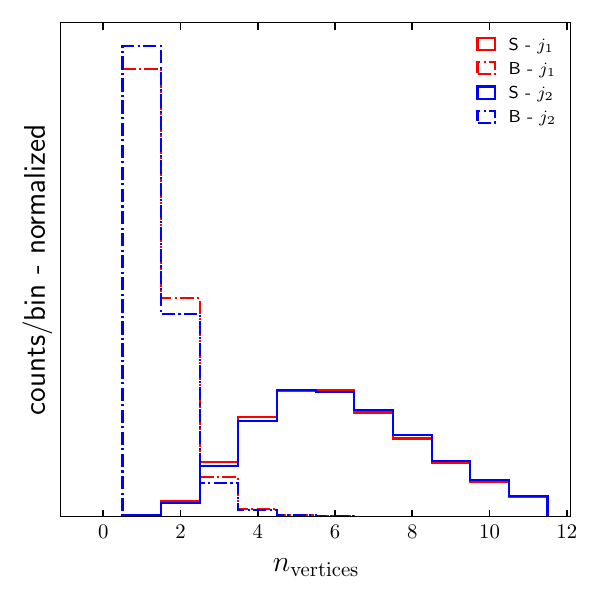}{$(m_{\pi'},  c\tau_{\pi'}) = (5~\mathrm{GeV},\ 0.3~\mathrm{mm})$}
\FloatBarrier
\clearpage

\subsection{\texorpdfstring{
		$(m_{\pi'},  c\tau_{\pi'}) = (10~\mathrm{GeV},\ 0.1~\mathrm{mm})$
	}
	{
		dark pion mass 10 GeV, dark pion lifetime 0.1 mm
	}
}
\subplotfeats{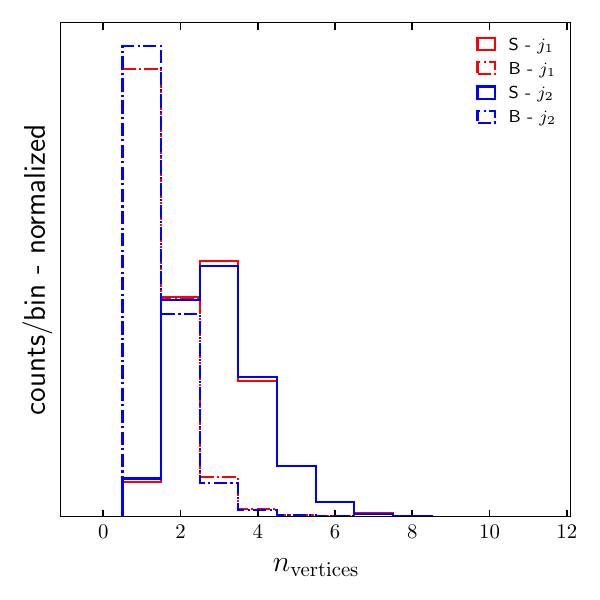}{$(m_{\pi'},  c\tau_{\pi'}) = (10~\mathrm{GeV},\ 0.1~\mathrm{mm})$}
\FloatBarrier
\clearpage

\subsection{\texorpdfstring{
		$(m_{\pi'},  c\tau_{\pi'}) = (10~\mathrm{GeV},\ 0.2~\mathrm{mm})$
	}
	{
		dark pion mass 10 GeV, dark pion lifetime 0.2 mm
	}
}
\subplotfeats{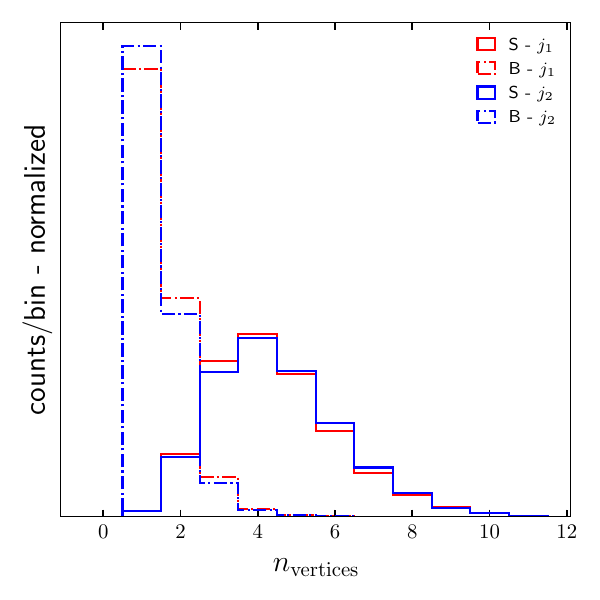}{$(m_{\pi'},  c\tau_{\pi'}) = (10~\mathrm{GeV},\ 0.2~\mathrm{mm})$}
\FloatBarrier
\clearpage

\subsection{\texorpdfstring{
		$(m_{\pi'},  c\tau_{\pi'}) = (10~\mathrm{GeV},\ 0.3~\mathrm{mm})$
	}
	{
		dark pion mass 10 GeV, dark pion lifetime 0.3 mm
	}
}
\subplotfeats{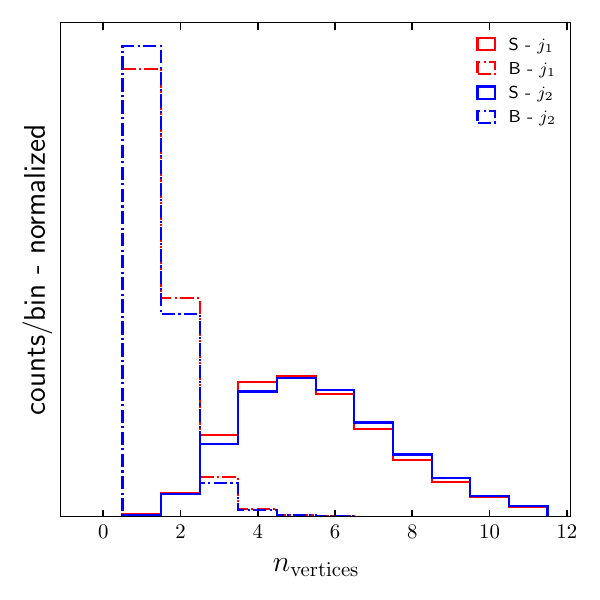}{$(m_{\pi'},  c\tau_{\pi'}) = (10~\mathrm{GeV},\ 0.3~\mathrm{mm})$}
\FloatBarrier
\clearpage

\section{Fit procedure}
\label{fit_procedure}

The sidebands were fit using \texttt{scipy.curve\_fit}, \texttt{scipy}'s~\cite{2020SciPy-NMeth} implementation of nonlinear least squares fit. The fit optimizes the cost function $L = \mathbf{r}^T \mathbf{r}$ where $\mathbf{r}_i$ is the residual in the $i$'th bin divided by the uncertainty in measured bin counts. The bin count uncertainty in a bin with $n$ counts was taken to be $\sqrt{n}$ according to Poisson statistics. The statistical uncertainty of expected counts in the signal region was estimated according to
\begin{equation}
	\begin{split}
		\sigma^2_{\mathrm{exp}} =
		\operatorname{Var}\left(
		\sum_{x \in \mathrm{sig. reg.}} N(x, \mathbf{p})
		\right) 
		\approx  \operatorname{Var}\left(
		\sum_{x \in \mathrm{sig. reg.}} \frac{d{N}}{d\mathbf{p}}(x, \hat{\mathbf{p}}) \cdot (\mathbf{p} - \hat{\mathbf{p}}) 
		\right) \\
		= 
		\left(\sum_{x \in \mathrm{sig. reg.}} \frac{d{N}}{d\mathbf{p}} \right)^T
		\operatorname{\mathbf{Cov}}
		\left(\sum_{x \in \mathrm{sig. reg.}} \frac{d{N}}{d\mathbf{p}} \right),
	\end{split}
\end{equation}
where $N(x, \mathbf{p})$ is the fit function from eq.~\eqref{fit_func} multiplied by the bin size, $\mathbf{p}$ is a random variable vector of fit function parameters, and $\mathbf{\hat{p}}$ are the estimated parameters. The covariance matrix for the parameters is estimated by
\begin{equation}
	\operatorname{\mathbf{Cov}} = \frac{L}{m-n} (\mathbf{J}^T \mathbf{J})^{-1},
\end{equation}
where $L$ is the cost function at $\mathbf{\hat{p}}$, $m$ is the number of points used for the fit, $n$ is the number of parameters ($=3$), and $\mathbf{J}$ is the Jacobian of $\mathbf{r}$ with respect to the parameters, evaluated at $\mathbf{\hat{p}}$.

\bibliographystyle{utphys}
\bibliography{ref}

\providecommand{\href}[2]{#2}\begingroup\raggedright\begin{thebibliography}{10}

\bibitem{Strassler:2006im}
M.~J. Strassler and K.~M. Zurek, ``{Echoes of a hidden valley at hadron
  colliders},'' \href{http://dx.doi.org/10.1016/j.physletb.2007.06.055}{{\em
  Phys. Lett. B} {\bfseries 651} (2007) 374--379},
  \href{http://arxiv.org/abs/hep-ph/0604261}{{\ttfamily arXiv:hep-ph/0604261}}.

\bibitem{Bai:2013xga}
Y.~Bai and P.~Schwaller, ``{Scale of dark QCD},''
  \href{http://dx.doi.org/10.1103/PhysRevD.89.063522}{{\em Phys. Rev. D}
  {\bfseries 89} no.~6, (2014) 063522},
  \href{http://arxiv.org/abs/1306.4676}{{\ttfamily arXiv:1306.4676 [hep-ph]}}.

\bibitem{Bernreuther:2019pfb}
E.~Bernreuther, F.~Kahlhoefer, M.~Krämer, and P.~Tunney, ``{Strongly
  interacting dark sectors in the early Universe and at the LHC through a
  simplified portal},'' \href{http://dx.doi.org/10.1007/JHEP01(2020)162}{{\em
  JHEP} {\bfseries 01} (2020) 162},
  \href{http://arxiv.org/abs/1907.04346}{{\ttfamily arXiv:1907.04346
  [hep-ph]}}.

\bibitem{Beauchesne:2018myj}
H.~Beauchesne, E.~Bertuzzo, and G.~Grilli Di~Cortona, ``{Dark matter in Hidden
  Valley models with stable and unstable light dark mesons},''
  \href{http://dx.doi.org/10.1007/JHEP04(2019)118}{{\em JHEP} {\bfseries 04}
  (2019) 118}, \href{http://arxiv.org/abs/1809.10152}{{\ttfamily
  arXiv:1809.10152 [hep-ph]}}.

\bibitem{Beauchesne:2019ato}
H.~Beauchesne and G.~Grilli~di Cortona, ``{Classification of dark pion
  multiplets as dark matter candidates and collider phenomenology},''
  \href{http://dx.doi.org/10.1007/JHEP02(2020)196}{{\em JHEP} {\bfseries 02}
  (2020) 196}, \href{http://arxiv.org/abs/1910.10724}{{\ttfamily
  arXiv:1910.10724 [hep-ph]}}.

\bibitem{Albouy:2022cin}
G.~Albouy {\em et~al.}, ``{Theory, phenomenology, and experimental avenues for
  dark showers: a Snowmass 2021 report},''
  \href{http://dx.doi.org/10.1140/epjc/s10052-022-11048-8}{{\em Eur. Phys. J.
  C} {\bfseries 82} no.~12, (2022) 1132},
  \href{http://arxiv.org/abs/2203.09503}{{\ttfamily arXiv:2203.09503
  [hep-ph]}}.

\bibitem{Cohen:2015toa}
T.~Cohen, M.~Lisanti, and H.~K. Lou, ``{Semivisible Jets: Dark Matter
  Undercover at the LHC},''
  \href{http://dx.doi.org/10.1103/PhysRevLett.115.171804}{{\em Phys. Rev.
  Lett.} {\bfseries 115} no.~17, (2015) 171804},
  \href{http://arxiv.org/abs/1503.00009}{{\ttfamily arXiv:1503.00009
  [hep-ph]}}.

\bibitem{Cohen:2017pzm}
T.~Cohen, M.~Lisanti, H.~K. Lou, and S.~Mishra-Sharma, ``{LHC Searches for Dark
  Sector Showers},'' \href{http://dx.doi.org/10.1007/JHEP11(2017)196}{{\em
  JHEP} {\bfseries 11} (2017) 196},
  \href{http://arxiv.org/abs/1707.05326}{{\ttfamily arXiv:1707.05326
  [hep-ph]}}.

\bibitem{Park:2017rfb}
M.~Park and M.~Zhang, ``{Tagging a jet from a dark sector with
  jet-substructures at colliders},''
  \href{http://dx.doi.org/10.1103/PhysRevD.100.115009}{{\em Phys. Rev. D}
  {\bfseries 100} no.~11, (2019) 115009},
  \href{http://arxiv.org/abs/1712.09279}{{\ttfamily arXiv:1712.09279
  [hep-ph]}}.

\bibitem{Cohen:2020afv}
T.~Cohen, J.~Doss, and M.~Freytsis, ``{Jet Substructure from Dark Sector
  Showers},'' \href{http://dx.doi.org/10.1007/JHEP09(2020)118}{{\em JHEP}
  {\bfseries 09} (2020) 118}, \href{http://arxiv.org/abs/2004.00631}{{\ttfamily
  arXiv:2004.00631 [hep-ph]}}.

\bibitem{Kar:2020bws}
D.~Kar and S.~Sinha, ``{Exploring jet substructure in semi-visible jets},''
  \href{http://dx.doi.org/10.21468/SciPostPhys.10.4.084}{{\em SciPost Phys.}
  {\bfseries 10} no.~4, (2021) 084},
  \href{http://arxiv.org/abs/2007.11597}{{\ttfamily arXiv:2007.11597
  [hep-ph]}}.

\bibitem{CMS:2021dzg}
{ CMS} Collaboration, A.~Tumasyan {\em et~al.}, ``{Search for resonant
  production of strongly coupled dark matter in proton-proton collisions at
  13~TeV},'' \href{http://dx.doi.org/10.1007/JHEP06(2022)156}{{\em JHEP}
  {\bfseries 06} (2022) 156}, \href{http://arxiv.org/abs/2112.11125}{{\ttfamily
  arXiv:2112.11125 [hep-ex]}}.

\bibitem{ATLAS:2023swa}
{ ATLAS} Collaboration, G.~Aad {\em et~al.}, ``{Search for non-resonant
  production of semi-visible jets using Run~2 data in ATLAS},''
  \href{http://arxiv.org/abs/2305.18037}{{\ttfamily arXiv:2305.18037
  [hep-ex]}}.

\bibitem{graphnet}
E.~Bernreuther, T.~Finke, F.~Kahlhoefer, M.~Kr\"amer, and A.~M\"uck, ``{Casting
  a graph net to catch dark showers},''
  \href{http://dx.doi.org/10.21468/SciPostPhys.10.2.046}{{\em SciPost Phys.}
  {\bfseries 10} no.~2, (2021) 046},
  \href{http://arxiv.org/abs/2006.08639}{{\ttfamily arXiv:2006.08639
  [hep-ph]}}.

\bibitem{Lu:2023gjk}
C.-T. Lu, H.~Lv, W.~Shen, L.~Wu, and J.~Zhang, ``{Probing Dark QCD Sector
  through the Higgs Portal with Machine Learning at the LHC},''
  \href{http://arxiv.org/abs/2304.03237}{{\ttfamily arXiv:2304.03237
  [hep-ph]}}.

\bibitem{Finke:2022lsu}
T.~Finke, M.~Kr\"amer, M.~Lipp, and A.~M\"uck, ``{Boosting mono-jet searches
  with model-agnostic machine learning},''
  \href{http://dx.doi.org/10.1007/JHEP08(2022)015}{{\em JHEP} {\bfseries 08}
  (2022) 015}, \href{http://arxiv.org/abs/2204.11889}{{\ttfamily
  arXiv:2204.11889 [hep-ph]}}.

\bibitem{Canelli:2021aps}
F.~Canelli, A.~de~Cosa, L.~L. Pottier, J.~Niedziela, K.~Pedro, and M.~Pierini,
  ``{Autoencoders for semivisible jet detection},''
  \href{http://dx.doi.org/10.1007/JHEP02(2022)074}{{\em JHEP} {\bfseries 02}
  (2022) 074}, \href{http://arxiv.org/abs/2112.02864}{{\ttfamily
  arXiv:2112.02864 [hep-ph]}}.

\bibitem{Schwaller:2015gea}
P.~Schwaller, D.~Stolarski, and A.~Weiler, ``{Emerging Jets},''
  \href{http://dx.doi.org/10.1007/JHEP05(2015)059}{{\em JHEP} {\bfseries 05}
  (2015) 059}, \href{http://arxiv.org/abs/1502.05409}{{\ttfamily
  arXiv:1502.05409 [hep-ph]}}.

\bibitem{CMS:2018bvr}
{ CMS} Collaboration, A.~M. Sirunyan {\em et~al.}, ``{Search for new particles
  decaying to a jet and an emerging jet},''
  \href{http://dx.doi.org/10.1007/JHEP02(2019)179}{{\em JHEP} {\bfseries 02}
  (2019) 179}, \href{http://arxiv.org/abs/1810.10069}{{\ttfamily
  arXiv:1810.10069 [hep-ex]}}.

\bibitem{AE1}
M.~Farina, Y.~Nakai, and D.~Shih, ``{Searching for New Physics with Deep
  Autoencoders},'' \href{http://dx.doi.org/10.1103/PhysRevD.101.075021}{{\em
  Phys. Rev. D} {\bfseries 101} no.~7, (2020) 075021},
  \href{http://arxiv.org/abs/1808.08992}{{\ttfamily arXiv:1808.08992
  [hep-ph]}}.

\bibitem{AE4}
T.~Heimel, G.~Kasieczka, T.~Plehn, and J.~M. Thompson, ``{QCD or What?},''
  \href{http://dx.doi.org/10.21468/SciPostPhys.6.3.030}{{\em SciPost Phys.}
  {\bfseries 6} no.~3, (2019) 030},
  \href{http://arxiv.org/abs/1808.08979}{{\ttfamily arXiv:1808.08979
  [hep-ph]}}.

\bibitem{AE2}
T.~Cheng, J.-F. Arguin, J.~Leissner-Martin, J.~Pilette, and T.~Golling,
  ``{Variational autoencoders for anomalous jet tagging},''
  \href{http://dx.doi.org/10.1103/PhysRevD.107.016002}{{\em Phys. Rev. D}
  {\bfseries 107} no.~1, (2023) 016002},
  \href{http://arxiv.org/abs/2007.01850}{{\ttfamily arXiv:2007.01850
  [hep-ph]}}.

\bibitem{AE3}
T.~Finke, M.~Kr\"amer, A.~Morandini, A.~M\"uck, and I.~Oleksiyuk,
  ``{Autoencoders for unsupervised anomaly detection in high energy physics},''
  \href{http://dx.doi.org/10.1007/JHEP06(2021)161}{{\em JHEP} {\bfseries 06}
  (2021) 161}, \href{http://arxiv.org/abs/2104.09051}{{\ttfamily
  arXiv:2104.09051 [hep-ph]}}.

\bibitem{Batson:2021agz}
J.~Batson, C.~G. Haaf, Y.~Kahn, and D.~A. Roberts, ``{Topological Obstructions
  to Autoencoding},'' \href{http://dx.doi.org/10.1007/JHEP04(2021)280}{{\em
  JHEP} {\bfseries 04} (2021) 280},
  \href{http://arxiv.org/abs/2102.08380}{{\ttfamily arXiv:2102.08380
  [hep-ph]}}.

\bibitem{Collins:2021nxn}
J.~H. Collins, P.~Mart\'\i{}n-Ramiro, B.~Nachman, and D.~Shih, ``{Comparing
  weak- and unsupervised methods for resonant anomaly detection},''
  \href{http://dx.doi.org/10.1140/epjc/s10052-021-09389-x}{{\em Eur. Phys. J.
  C} {\bfseries 81} no.~7, (2021) 617},
  \href{http://arxiv.org/abs/2104.02092}{{\ttfamily arXiv:2104.02092
  [hep-ph]}}.

\bibitem{yu2015learning}
F.~X. Yu, K.~Choromanski, S.~Kumar, T.~Jebara, and S.-F. Chang, ``On learning
  from label proportions,'' \href{http://arxiv.org/abs/1402.5902}{{\ttfamily
  arXiv:1402.5902 [stat.ML]}}.

\bibitem{JMLR:v10:quadrianto09a}
N.~Quadrianto, A.~J. Smola, T.~S. Caetano, and Q.~V. Le, ``Estimating labels
  from label proportions,'' {\em Journal of Machine Learning Research}
  {\bfseries 10} no.~82, (2009) 2349--2374.
  \url{http://jmlr.org/papers/v10/quadrianto09a.html}.

\bibitem{CWoLa1}
L.~M. Dery, B.~Nachman, F.~Rubbo, and A.~Schwartzman, ``{Weakly Supervised
  Classification in High Energy Physics},''
  \href{http://dx.doi.org/10.1007/JHEP05(2017)145}{{\em JHEP} {\bfseries 05}
  (2017) 145}, \href{http://arxiv.org/abs/1702.00414}{{\ttfamily
  arXiv:1702.00414 [hep-ph]}}.

\bibitem{Cohen:2017exh}
T.~Cohen, M.~Freytsis, and B.~Ostdiek, ``{(Machine) Learning to Do More with
  Less},'' \href{http://dx.doi.org/10.1007/JHEP02(2018)034}{{\em JHEP}
  {\bfseries 02} (2018) 034}, \href{http://arxiv.org/abs/1706.09451}{{\ttfamily
  arXiv:1706.09451 [hep-ph]}}.

\bibitem{CWoLa2}
E.~M. Metodiev, B.~Nachman, and J.~Thaler, ``{Classification without labels:
  Learning from mixed samples in high energy physics},''
  \href{http://dx.doi.org/10.1007/JHEP10(2017)174}{{\em JHEP} {\bfseries 10}
  (2017) 174}, \href{http://arxiv.org/abs/1708.02949}{{\ttfamily
  arXiv:1708.02949 [hep-ph]}}.

\bibitem{CWoLa3}
P.~T. Komiske, E.~M. Metodiev, B.~Nachman, and M.~D. Schwartz, ``{Learning to
  classify from impure samples with high-dimensional data},''
  \href{http://dx.doi.org/10.1103/PhysRevD.98.011502}{{\em Phys. Rev. D}
  {\bfseries 98} no.~1, (2018) 011502(R)},
  \href{http://arxiv.org/abs/1801.10158}{{\ttfamily arXiv:1801.10158
  [hep-ph]}}.

\bibitem{CWoLa4}
J.~H. Collins, K.~Howe, and B.~Nachman, ``{Extending the search for new
  resonances with machine learning},''
  \href{http://dx.doi.org/10.1103/PhysRevD.99.014038}{{\em Phys. Rev. D}
  {\bfseries 99} no.~1, (2019) 014038},
  \href{http://arxiv.org/abs/1902.02634}{{\ttfamily arXiv:1902.02634
  [hep-ph]}}.

\bibitem{ATLAS:2020iwa}
{ ATLAS} Collaboration, G.~Aad {\em et~al.}, ``{Dijet resonance search with
  weak supervision using $\sqrt{s}=13$~TeV $pp$ collisions in the ATLAS
  detector},'' \href{http://dx.doi.org/10.1103/PhysRevLett.125.131801}{{\em
  Phys. Rev. Lett.} {\bfseries 125} no.~13, (2020) 131801},
  \href{http://arxiv.org/abs/2005.02983}{{\ttfamily arXiv:2005.02983
  [hep-ex]}}.

\bibitem{Amram:2020ykb}
O.~Amram and C.~M. Suarez, ``{Tag N\textquoteright{} Train: a technique to
  train improved classifiers on unlabeled data},''
  \href{http://dx.doi.org/10.1007/JHEP01(2021)153}{{\em JHEP} {\bfseries 01}
  (2021) 153}, \href{http://arxiv.org/abs/2002.12376}{{\ttfamily
  arXiv:2002.12376 [hep-ph]}}.

\bibitem{Choudalakis:2011qn}
G.~Choudalakis, ``{On hypothesis testing, trials factor, hypertests and the
  BumpHunter},'' in {\em {PHYSTAT 2011}}.
\newblock 1, 2011.
\newblock \href{http://arxiv.org/abs/1101.0390}{{\ttfamily arXiv:1101.0390
  [physics.data-an]}}.

\bibitem{ATLAS:2019fgd}
{ ATLAS} Collaboration, G.~Aad {\em et~al.}, ``{Search for new resonances in
  mass distributions of jet pairs using 139~fb$^{-1}$ of $pp$ collisions at
  $\sqrt{s}=13$~TeV with the ATLAS detector},''
  \href{http://dx.doi.org/10.1007/JHEP03(2020)145}{{\em JHEP} {\bfseries 03}
  (2020) 145}, \href{http://arxiv.org/abs/1910.08447}{{\ttfamily
  arXiv:1910.08447 [hep-ex]}}.

\bibitem{ATLAS:2012mwf}
{ ATLAS} Collaboration, G.~Aad {\em et~al.}, ``{Measurement of the flavour
  composition of dijet events in $pp$ collisions at $\sqrt{s}=7$~TeV with the
  ATLAS detector},''
  \href{http://dx.doi.org/10.1140/epjc/s10052-013-2301-5}{{\em Eur. Phys. J. C}
  {\bfseries 73} no.~2, (2013) 2301},
  \href{http://arxiv.org/abs/1210.0441}{{\ttfamily arXiv:1210.0441 [hep-ex]}}.

\bibitem{Carloni:2010tw}
L.~Carloni and T.~Sjostrand, ``{Visible Effects of Invisible Hidden Valley
  Radiation},'' \href{http://dx.doi.org/10.1007/JHEP09(2010)105}{{\em JHEP}
  {\bfseries 09} (2010) 105}, \href{http://arxiv.org/abs/1006.2911}{{\ttfamily
  arXiv:1006.2911 [hep-ph]}}.

\bibitem{ATLAS:2017zuf}
{ ATLAS} Collaboration, M.~Aaboud {\em et~al.}, ``{Search for diboson
  resonances with boson-tagged jets in $pp$ collisions at $\sqrt{s}=13$~TeV
  with the ATLAS detector},''
  \href{http://dx.doi.org/10.1016/j.physletb.2017.12.011}{{\em Phys. Lett. B}
  {\bfseries 777} (2018) 91--113},
  \href{http://arxiv.org/abs/1708.04445}{{\ttfamily arXiv:1708.04445
  [hep-ex]}}.

\bibitem{CMS:2016rqm}
{ CMS} Collaboration, A.~M. Sirunyan {\em et~al.}, ``{Search for massive
  resonances decaying into WW, WZ or ZZ bosons in proton-proton collisions at
  $\sqrt{s} = 13$~TeV},'' \href{http://dx.doi.org/10.1007/JHEP03(2017)162}{{\em
  JHEP} {\bfseries 03} (2017) 162},
  \href{http://arxiv.org/abs/1612.09159}{{\ttfamily arXiv:1612.09159
  [hep-ex]}}.

\bibitem{Benkendorfer:2020gek}
K.~Benkendorfer, L.~{Le Pottier}, and B.~Nachman, ``{Simulation-assisted
  decorrelation for resonant anomaly detection},''
  \href{http://dx.doi.org/10.1103/PhysRevD.104.035003}{{\em Phys. Rev. D}
  {\bfseries 104} no.~3, (2021) 035003},
  \href{http://arxiv.org/abs/2009.02205}{{\ttfamily arXiv:2009.02205
  [hep-ph]}}.

\bibitem{Alwall:2014hca}
J.~Alwall, R.~Frederix, S.~Frixione, V.~Hirschi, F.~Maltoni, O.~Mattelaer,
  H.~S. Shao, T.~Stelzer, P.~Torrielli, and M.~Zaro, ``{The automated
  computation of tree-level and next-to-leading order differential cross
  sections, and their matching to parton shower simulations},''
  \href{http://dx.doi.org/10.1007/JHEP07(2014)079}{{\em JHEP} {\bfseries 07}
  (2014) 079}, \href{http://arxiv.org/abs/1405.0301}{{\ttfamily arXiv:1405.0301
  [hep-ph]}}.

\bibitem{Ball:2013hta}
{ NNPDF} Collaboration, R.~D. Ball, V.~Bertone, S.~Carrazza, L.~Del~Debbio,
  S.~Forte, A.~Guffanti, N.~P. Hartland, and J.~Rojo, ``{Parton distributions
  with QED corrections},''
  \href{http://dx.doi.org/10.1016/j.nuclphysb.2013.10.010}{{\em Nucl. Phys. B}
  {\bfseries 877} (2013) 290--320},
  \href{http://arxiv.org/abs/1308.0598}{{\ttfamily arXiv:1308.0598 [hep-ph]}}.

\bibitem{Sjostrand:2014zea}
T.~Sj\"ostrand, S.~Ask, J.~R. Christiansen, R.~Corke, N.~Desai, P.~Ilten,
  S.~Mrenna, S.~Prestel, C.~O. Rasmussen, and P.~Z. Skands, ``{An introduction
  to PYTHIA 8.2},'' \href{http://dx.doi.org/10.1016/j.cpc.2015.01.024}{{\em
  Comput. Phys. Commun.} {\bfseries 191} (2015) 159--177},
  \href{http://arxiv.org/abs/1410.3012}{{\ttfamily arXiv:1410.3012 [hep-ph]}}.

\bibitem{deFavereau:2013fsa}
{ DELPHES 3} Collaboration, J.~de~Favereau, C.~Delaere, P.~Demin, A.~Giammanco,
  V.~Lema\^\i{}tre, A.~Mertens, and M.~Selvaggi, ``{DELPHES 3, A modular
  framework for fast simulation of a generic collider experiment},''
  \href{http://dx.doi.org/10.1007/JHEP02(2014)057}{{\em JHEP} {\bfseries 02}
  (2014) 057}, \href{http://arxiv.org/abs/1307.6346}{{\ttfamily arXiv:1307.6346
  [hep-ex]}}.

\bibitem{CMS:2014pgm}
{ CMS} Collaboration, S.~Chatrchyan {\em et~al.}, ``{Description and
  performance of track and primary-vertex reconstruction with the CMS
  tracker},'' \href{http://dx.doi.org/10.1088/1748-0221/9/10/P10009}{{\em
  JINST} {\bfseries 9} no.~10, (2014) P10009},
  \href{http://arxiv.org/abs/1405.6569}{{\ttfamily arXiv:1405.6569
  [physics.ins-det]}}.

\bibitem{Cacciari:2008gp}
M.~Cacciari, G.~P. Salam, and G.~Soyez, ``{The anti-$k_t$ jet clustering
  algorithm},'' \href{http://dx.doi.org/10.1088/1126-6708/2008/04/063}{{\em
  JHEP} {\bfseries 04} (2008) 063},
  \href{http://arxiv.org/abs/0802.1189}{{\ttfamily arXiv:0802.1189 [hep-ph]}}.

\bibitem{Fruhwirth:2007hz}
R.~Fruhwirth, W.~Waltenberger, and P.~Vanlaer, ``{Adaptive vertex fitting},''
  \href{http://dx.doi.org/10.1088/0954-3899/34/12/N01}{{\em J. Phys. G}
  {\bfseries 34} (2007) N343}.

\bibitem{5734880}
W.~{Waltenberger}, ``Rave—a detector-independent toolkit to reconstruct
  vertices,'' \href{http://dx.doi.org/10.1109/TNS.2011.2119492}{{\em IEEE
  Transactions on Nuclear Science} {\bfseries 58} no.~2, (2011) 434--444}.
  \url{https://rave.hepforge.org/}.

\bibitem{chollet2015keras}
F.~Chollet {\em et~al.}, ``Keras.'' \url{https://keras.io}, 2015.

\bibitem{tensorflow2015-whitepaper}
M.~Abadi {\em et~al.}, ``{TensorFlow}: Large-scale machine learning on
  heterogeneous systems,'' 2015.
\newblock \url{https://www.tensorflow.org/}.

\bibitem{2020SciPy-NMeth}
P.~Virtanen {\em et~al.}, ``{{SciPy} 1.0: Fundamental Algorithms for Scientific
  Computing in Python},''
  \href{http://dx.doi.org/10.1038/s41592-019-0686-2}{{\em Nature Methods}
  {\bfseries 17} (2020) 261--272}.

\end{thebibliography}\endgroup

\end{document}